\documentclass[reprint,amsmath,amssymb,aps,prb]{revtex4-1}

\usepackage[caption = false, labelformat = empty, farskip = -22pt]{subfig}
\usepackage{graphicx}
\usepackage{dcolumn}
\usepackage{bm}
\usepackage{xcolor}
\usepackage{braket}
\usepackage{appendix}
\usepackage{amsmath}
\usepackage{comment}
\usepackage[english]{babel}

\usepackage{xcolor}

\makeatletter
\def\mathcolor#1#{\@mathcolor{#1}}
\def\@mathcolor#1#2#3{%
  \protect\leavevmode
  \begingroup
    \color#1{#2}#3%
  \endgroup
}

\begin{document}

\title{Superconducting proximity effect and order parameter fluctuations in disordered and quasiperiodic systems}

\author{Gautam Rai}
\email{gautamra@usc.edu}

\author{Stephan Haas}%

\affiliation{%
Department of Physics and Astronomy, University of Southern California, Los Angeles, California 90089-0484, USA
}%

\author{Anuradha Jagannathan}
\affiliation{
Laboratoire de Physique des Solides, Universit\'{e} Paris-Sud, 91405 Orsay, France
}%

\date{\today}

    \begin{abstract}
    We study the superconducting proximity effect in inhomogeneous systems in which a disordered or quasicrystalline normal-state wire is connected to a BCS superconductor. We self-consistently compute the local superconducting order parameters in the real space Bogoliubov-de Gennes framework for three cases, namely, when states are i) extended, ii) localized or iii) critical. The results show that the spatial decay of the superconducting order parameter as one moves away from the normal-superconductor interface is power law in cases i) and iii), stretched exponential in case ii). In the quasicrystalline case, we observe self-similarity in the spatial modulation of the proximity-induced superconducting order parameter. To characterize fluctuations, which are large in these systems, we study the distribution functions of the order parameter at the center of the normal region. These are Gaussian functions of the variable (case i)  or of its logarithm (cases ii and iii). We give arguments to explain the characteristics of the distributions and their scaling with system size for each of the three cases.  
    \end{abstract}
    
    \maketitle
    
    \section{Introduction}
    
   In this paper we study the superconducting pairing correlations induced by the proximity effect \cite{de_gennes_boundary_1964} in inhomogeneous systems.  We will consider, specifically, disordered and quasiperiodic chains, whose wave functions are not described by the Bloch theorem. While the subject of bulk disordered superconductors has been addressed by many previous theoretical \cite{ghosal_role_1998,feigelman_eigenfunction_2007} and recent experimental \cite{dubouchet_collective_2019, sacepe_localization_2011} works, little is known about the real space dependence and the nature of the fluctuations in the proximity-induced superconductivity in inhomogeneous systems. The aim of the present paper is to provide a detailed picture of the spatial decay and the fluctuations of the proximity-induced superconducting order parameter (OP) in such inhomogeneous systems.

We recall that the proximity effect provides a useful probe of electronic properties. A number of recent works have considered the superconducting proximity effect in  a variety of systems: topological insulators \cite{chiu_induced_2016, chevallier_andreev_2013}, monolayer and bilayer graphene \cite{ojeda-aristizabal_tuning_2009, titov_josephson_2006, black-schaffer_self-consistent_2008}, interacting chains \cite{affleck_andreev_2000} and one-dimensional quasicrystals \cite{rai_proximity_2019}. 

    We will consider two archetypal 1D systems: the Anderson model \cite{anderson_localized_1961} and the Fibonacci hopping model \cite{ostlund_one-dimensional_1983, kohmoto_localization_1983}, both of which  have been extensively studied since their introduction. 
    We use a real-space version of the Bogoliubov-de Gennes mean-field approach to determine the spatial dependence of the self-consistently calculated local superconducting order parameter (OP), $\Delta(x)$, for a set of samples---chains with a disordered/quasiperiodic configuration corresponding to a fixed pairing strength and a fixed length. We find that the ensemble-averaged OP decays as a power law when the wavefunctions are quasi-extended as in the weak-disorder limit of the Anderson model, or power-law localized as in the Fibonacci chain. On the other hand, it decays exponentially on the scale of the system size when the wavefunctions are strongly localized as in the strong-disorder regime of the Anderson model.
    
    The above statements hold for the mean (or typical) value. To more completely characterize the fluctuations around the mean, we compute the distribution functions of the values of OP for a given position in each of the three cases. These distribution functions reflect the nature of the electronic states in the systems, which are respectively quasi-extended states, strongly localized states, and critical (multi-fractal) states. We show that in the weak-disorder case, the proximity-induced superconducting order parameter is described by a Gaussian distribution, whereas it is described by log-normal distributions in  quasiperiodic and strongly disordered systems.
    
We close this introduction by mentioning some works on superconducting order in $bulk$ inhomogeneous systems for closely related models. Ghosal et al \cite{ghosal_inhomogeneous_2001} obtained some early results for the distribution of order parameters for bulk disordered superconductors. Their theoretical prediction of superconducting \emph{islands} in a non-superconducting \emph{sea} have led to STM based experiments on 2D films \cite{randeria_scanning_2016, sacepe_disorder-induced_2008}. Bulk superconductivity in 2D quasicrystals have been studied by real space DMFT \cite{sakai_superconductivity_2017} and also using a mean field analysis as we do here \cite{sakai_exotic_2019, araujo_conventional_2019}. Similar mean-field calculations have been used to study the behavior of the superconducting singlet and triplet order parameter near edges and impurities in 1D and 2D systems \cite{lauke_friedel_2018}. Induced pair correlations have been studied in detail in clean ferromagnet-superconductor heterostructures\cite{halterman_proximity_2001}, while prominent spectral features have been characterized in the diffusive counterpart \cite{alidoust_zero-energy_2015}. Few experimental investigations exist for proximity-induced superconductivity in disordered systems and, thus far, none in quasiperiodic systems. Disordered wires in the diffusive regime have been investigated in \cite{ gueron_superconducting_1996,le_sueur_phase_2008}, and disordered 2D films in \cite{serrier-garcia_scanning_2013}. It would be therefore interesting to carry out investigations to test the predictions of this paper for the three paradigmatic situations that we describe here. 

The paper is organized as follows: in Section \ref{sec:BdG} we present a review of the real-space Bogoliubov-de Gennes mean field approach that is used here in a self-consistent manner. In Section \ref{sec:Models} we discuss the Hamiltonian and the choice of parameters. In Section \ref{sec:Disordered} we analyze the results obtained for the Anderson model and in Section \ref{sec:Fibonacci} we analyze the results obtained for the Fibonacci model. Section \ref{sec:Conclusion} summarizes our conclusions.


    \begin{figure*}
    \subfloat[\label{fig:ring}]{
    \includegraphics[height=0.22\textwidth]{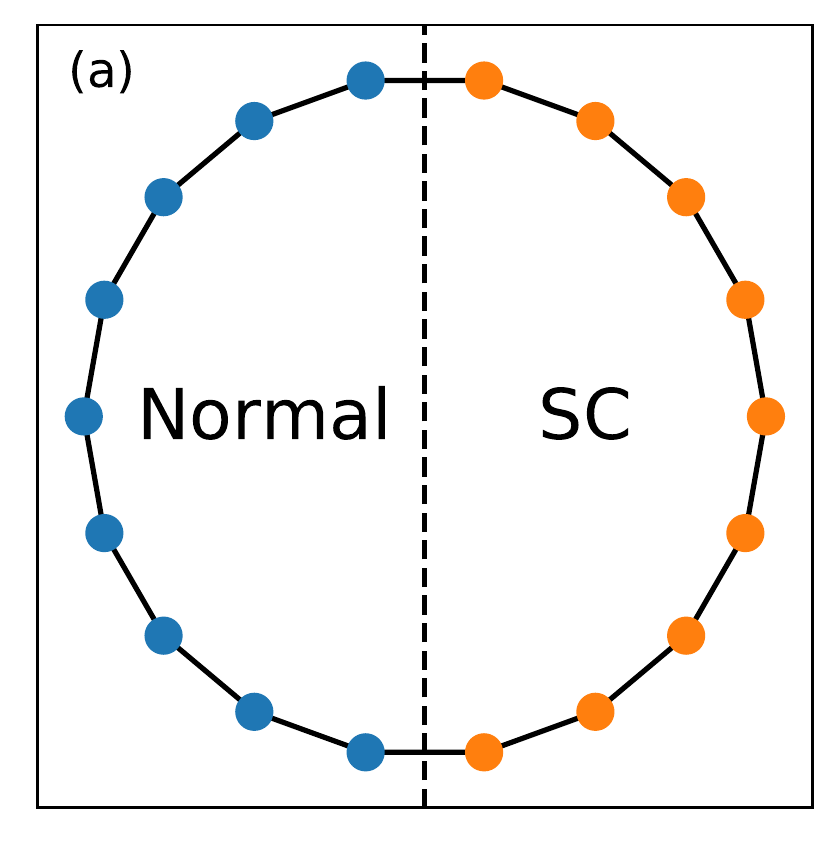}}
    \subfloat[\label{fig:per-Delta}]{
    \includegraphics[height=0.22\textwidth]{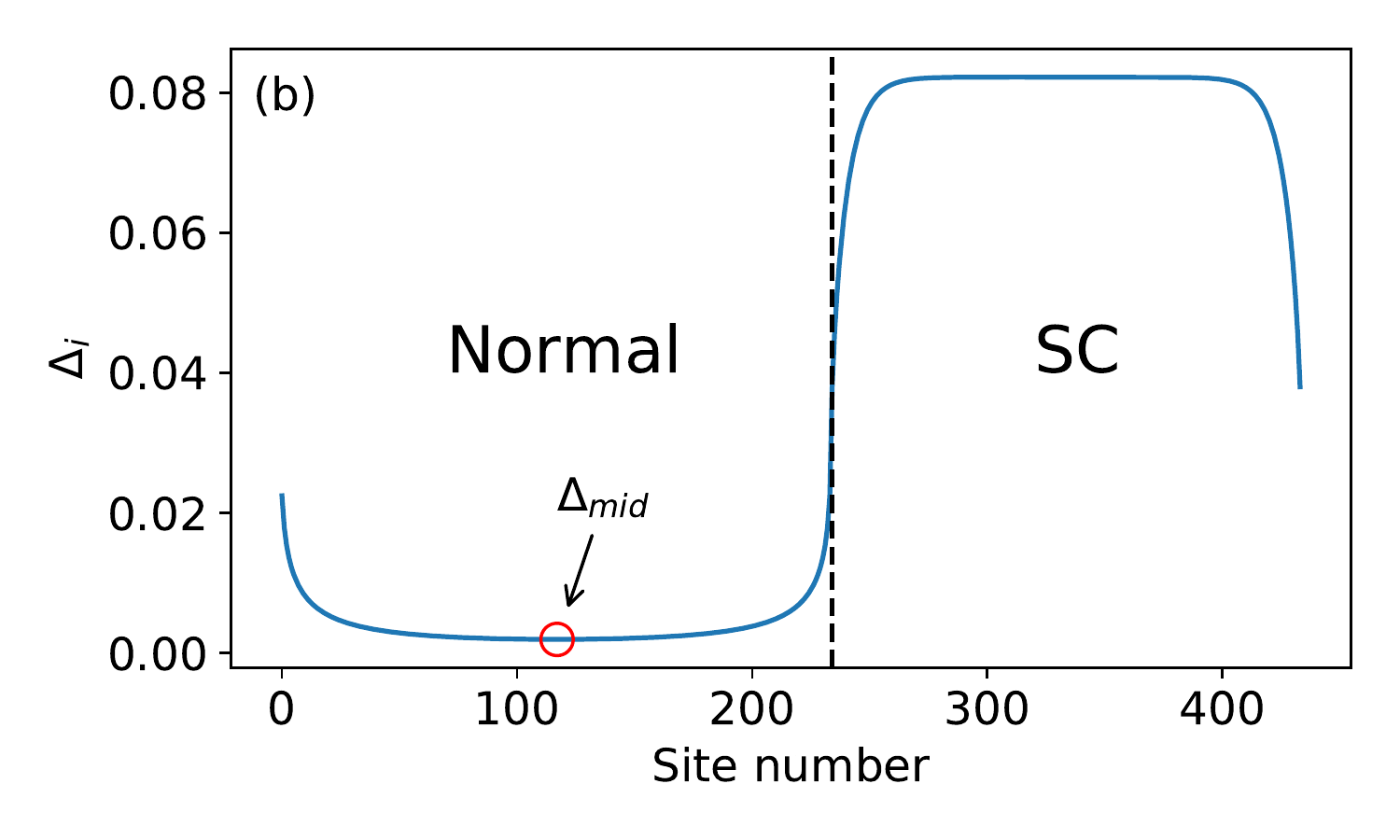}}
    \subfloat[\label{fig:per-scaling}]{
    \includegraphics[height=0.22\textwidth]{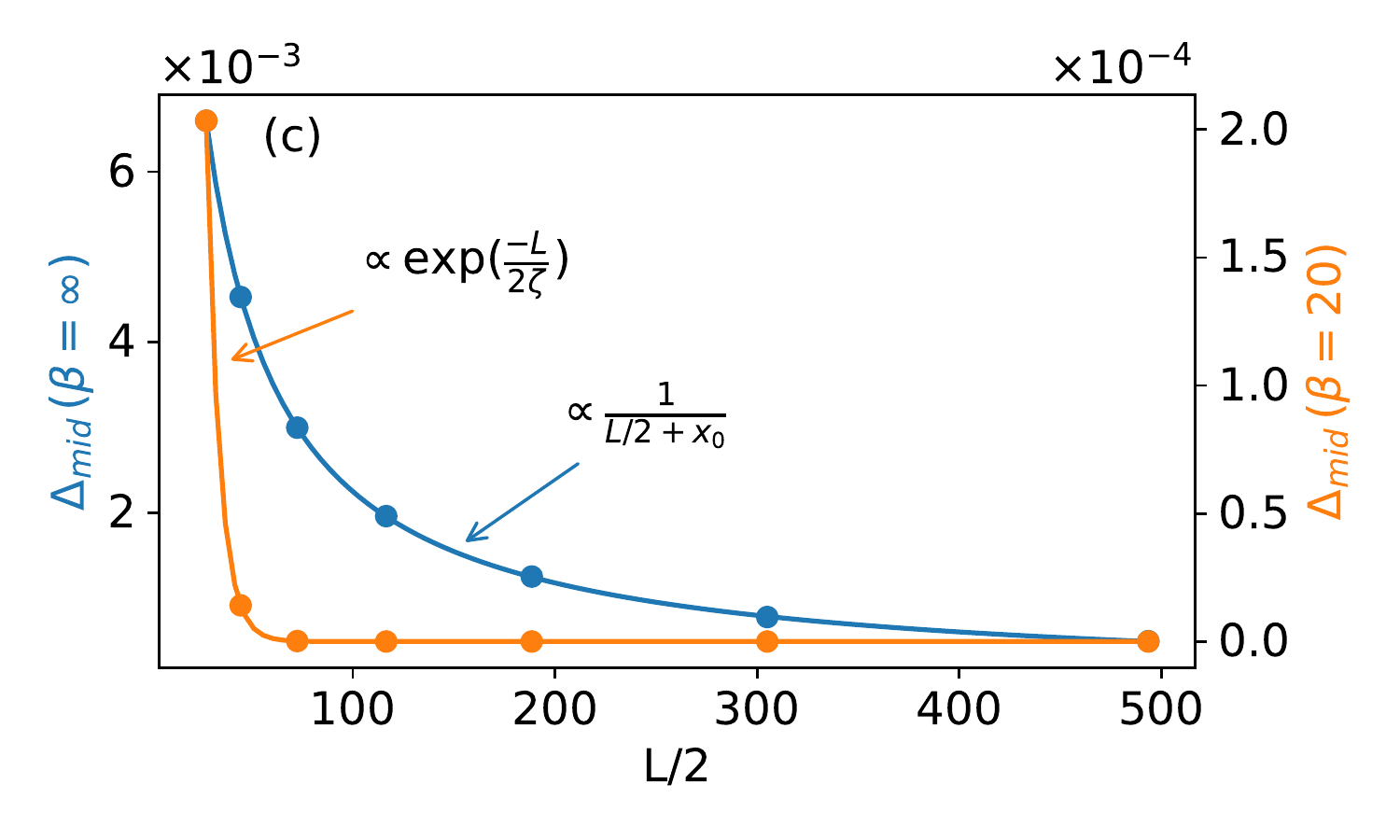}}
    
    \caption{Superconducting proximity effect in a normal one-dimensional metal: (a) Sketch of a normal-superconductor (N-SC) hybrid ring: on the left side is the non-superconducting region, whereas on the right is an intrinsic superconductor. (b) Spatial variation of the superconducting order parameter $\Delta_i$ along the chain for a clean N-SC hybrid ring---both parts are bulk-periodic. (c) Spatial decay of the superconducting order parameter in the normal part of a clean N-SC ring for zero and finite temperature ($\beta = 1/kT$), extracted by finite-size scaling of $\Delta_{mid}$ for chains with varying $L$. The best fit parameters are $x_o = 10.0$ and $\zeta = 6.4$. $\Delta$ decays inversely with distance at zero temperature and exponentially with distance at finite temperature.}
    \end{figure*}
    
    \section{The Bogoliubov-de Gennes method for inhomogeneous chains}\label{sec:BdG}
    
    The starting point for our discussion is the attractive Hubbard model \cite{de2018superconductivity},
    
    \begin{align}
        \hat{H} = \sum_i \Bigg\{&
        \sum_\alpha (\epsilon_i^0 - \mu_i) c_{i\alpha}^\dagger c_{i\alpha}\nonumber \\
        &+\sum_\alpha t_{i} c_{i\alpha}^\dagger c_{i+1\,\alpha}  + h.c.\nonumber\\
        &-\sum_{\alpha\beta} \frac{V_i}{2}c^\dagger_{i\alpha} c^\dagger_{i\beta} c_{i\beta} c_{i\alpha} \Bigg\}.
    \end{align}
    
    Here $c_{i\alpha}$ is the electron annihilation operator at site $i$ ($i=1,...,L_{tot}$, where $L_{tot}$ is the total number of sites) and with spin $\alpha$. $\epsilon_i^0$ and $\mu_i$ are the on-site potential and the chemical potential respectively at site $i$, $t_{i}$ is the hopping amplitude between sites $i$ and $i+1$ and $V_i>0$ is the strength of the attractive Hubbard interaction between spin-up and spin-down electrons at site $i$. In the Bogoliubov-de Gennes method, one introduces the mean fields $\epsilon^{HF}_i$ and $\Delta_i$ to decouple the quartic interaction term. The resulting mean field Hamiltonian is given by
    \begin{align}
        \hat{H}_{mf} = \sum_i \Bigg\{ &\sum_\alpha \overbrace{(\epsilon_i^0 + {\epsilon^{HF}_i} - \mu_i)}^{\epsilon_i} c^\dagger_{i\alpha} c_{i\alpha}\nonumber\\
        & + \sum_\alpha t_{i\,i+1} c^\dagger_{i\alpha} c_{i+1\,\alpha} + h.c.\nonumber\\
        & + V_i {\Delta_i} c_{i\uparrow}^\dagger c_{i\downarrow}^\dagger + h.c. \Bigg\}.\label{eq:mfham}
    \end{align}
    $\hat{H}_{mf}$ can be diagonalized by introducing the Bogoliubov transformation:
    
    \begin{align}
        c_{i\uparrow} &= \sum_n \gamma_{n\uparrow} u_{in} - \gamma^\dagger_{n\downarrow} v^*_{in},\\
        c_{i\downarrow} &= \sum_n \gamma_{n\downarrow} u_{in} + \gamma^\dagger_{n\uparrow} v^*_{in}.
    \end{align}
    
    One solves the resulting Bogoliubov equations in real space to obtain the eigenstates and their energies.  Introducing the $L_{tot}$-component vectors  $u_n$ and $v_n$, where $u_n$ (resp. $v_n$) are the the coefficients $u_{in}$  (resp. $v_{in})$ with $i=1,..,L_{tot}$, the matrix form of these equations reads
    \begin{align}
        \begin{pmatrix}
        \hat{K}  &  \hat{\Delta}\\
       \hat{\Delta}          &  -\hat{K}
        \end{pmatrix}
        \begin{pmatrix}
        u_n\\v_n
        \end{pmatrix} = 
        E_n\begin{pmatrix}
        u_n\\v_n
        \end{pmatrix},
        \label{eq:BdG}
    \end{align}
    where $E_n$ is the associated eigenvalue.
    The components of the matrices $\hat{K}$ and $\hat{\Delta}$ are given in terms of the Kronecker delta $\delta_{ij}$ by
    \begin{align}
        {\Delta}_{ij} &= \sum_{i} V_i \Delta_i {\delta}_{ij},\label{eq:MatDel}\\
        {K}_{ij}     &=\sum_{i} \epsilon_i {\delta}_{ij} + t_{i} {\delta}_{i+1\,j} + t_{i-1}{\delta}_{i-1\,j}.\label{eq:MatK}
    \end{align}
Due to the redundancy of this system of equations, it suffices to keep only the positive energy solutions $E_n>0$. The eigenvectors satisfy normalization conditions $\sum_n |u_{in}|^2 + |v_{in}|^2 = 1$ at each site $i$.   
    
    The mean-field Hamiltonian \eqref{eq:mfham} contains  $L_{tot}$ local superconducting order parameters $\Delta_i$ and $L_{tot}$ local effective onsite energies $\epsilon_i=\epsilon_i^0+\epsilon_i^{HF}-\mu_i$. These quantities must satisfy the self-consistency equations,
    \begin{align}
    \epsilon^{HF}_i &= -V\sum_n |u_{in}|^2 f(E_n, T) + |v_{in}|^2 (1 - f(E_n, T)),\nonumber\\
    \Delta_i &= \sum_n v^*_{in} u_{in} (1 - 2f(E_n, T)),\label{eq:self-con}
    \end{align}
    where $f(E_n, T)$ is the Fermi-Dirac distribution function at temperature $T$. In our numerical calculations, we begin with a starting ansatz for the $2L_{tot}$ mean-field parameters. We then iteratively solve the eigenvalue problem Eq.~\eqref{eq:BdG}, and compute the new values of these parameters using Eq.~\eqref{eq:self-con}. This procedure is repeated until convergence has been achieved to a reasonable accuracy \footnote{At every iteration step, the error $err_i = \Delta_i - \Delta^{old}_i$. Convergence is reached if $\Delta_i$ at every position with a non-zero $\Delta_i$ has an error of less than 0.1\%. Conversely, if for any position $i$, $err_i/\Delta_i > 0.001$ and $\Delta_i$ is appreciably different from zero ($\gg$ machine precision $10^{-16}$), then convergence is not reached.
}.
    
    Mean-field methods similar to this have been previously applied to the problem of inhomogeneous superconductors in \cite{sakai_exotic_2019, ghosal_inhomogeneous_2001, ghosal_role_1998, chiu_induced_2016, black-schaffer_self-consistent_2008, lauke_friedel_2018, cayssol_isolated_2003, rai_proximity_2019, araujo_conventional_2019}. Quantum fluctuations are generally too strong for mean-field treatments to sufficiently describe low-dimensional systems. There are however two contexts in which a one-dimensional mean-field description is meaningful: 1) In an experiment, a 1D quantum wire will be embedded in a 3D environment. Substrate and other surrounding media may conspire to subdue quantum fluctuations while the effective description for the quantum wire is 1D. 2) Quasi-1D systems such as nanotubes or mesoscopic wires often admit an effective 1D description, while the true dimensionality of the system is 3D and the mean-field results apply.
    
    \section{Description of Models}\label{sec:Models}
    
    We study hybrid chains, consisting of a normal, i.e. non-interacting, region (N) and a superconducting region (SC). Closed boundary conditions are assumed (see Fig. \ref{fig:ring}), such that the system has a ring geometry with two N-SC interfaces. Within the formalism outlined in the preceding section, this system can be fully described by appropriately specifying the matrices $\hat{\Delta}$, $\hat{K}$ in Eqs.~\eqref{eq:MatDel} and \eqref{eq:MatK}.  
            
    As our primary goal is to examine the effects of disorder on the proximity effect, we will focus on the ideal case where the interfaces are transparent. We will measure all energies in units of the strength of the nearest neighbor hopping in the superconductor, $t$. The hopping energy in the superconductor is then $t = -1$. The hopping amplitudes in the N region will be chosen to be of comparable strength, i.e. of the order of unity. In each of the models, the band fillings are fixed at $\frac{1}{2}$, so that the Fermi level is in the middle of the spectrum and particle-hole symmetry is maintained. The strength of the attractive Hubbard interaction in the SC is set to a fixed value \footnote{The BCS pairing attraction is chosen to be of the same order of magnitude as the nearest-neighbor hopping in order to resolve superconducting features in systems of this size.} for $ L \leq i\leq L_{tot}-1$.. The lengths of the SC region are chosen to be large enough that the OP relaxes to attain the expected bulk value well inside the SC region. The length of the N region is likewise chosen large enough, such that the bulk penetration laws can be properly determined by finite size scaling. 
    
    This amounts to the following set of choices for the parameters:
    
    \begin{itemize}
    \item{The normal region corresponds to the first $L$ sites, $i=0,..,L-1$. The superconducting region is of length $L_{SC}$, corresponding to indices $i=L,L+1,...,L_{tot}-1$ where $L_{tot}=L+L_{SC}$ is the total number of sites. $L_{SC} = 200$ everywhere, $L$ ranges from $90$ to $1598$.}
    \item{The hoppings at the two interfaces are taken to be unity, i.e. $t_{L-1} = t_{L_{tot}-1} = -1$.}
    \item{There are no interactions in the normal region, i.e. $V_i = 0$ for $0\leq i \leq L-1$.}
    \item{Within the N region, $t_i$ values are either sampled from a random distribution function (see Sec.IV) or taken from a Fibonacci sequence (see Sec.V)}
    \item{Within the (translationally invariant) superconductor , $t_{i} = -1, V_i/t = 1.5$}
    \item{Both regions are at half-filling, i.e. the Fermi level is in the middle of the spectrum. In the normal part, this means $\epsilon_i^0-\mu_i = 0$. In the superconducting part, we need to account for the Hartree-Fock shift which implies $\epsilon_i^0-\mu_i=-\epsilon_i^{HF} = \frac{V}{2}$.}
    \end{itemize}
   
    Throughout this study, the central  observable of interest is the strength of the superconducting order parameter at the mid-point of the normal region of the ring, $\Delta_{mid}$. This is given by $\Delta_{\frac{L -1}{2}}$ for odd chains and $\frac{1}{2}\left(\Delta_{\frac{L}{2}}+ \Delta_{\frac{L}{2} -1}\right)$ for even chains. We compute $\Delta_{mid}$ for an ensemble of rings with a given size and disorder/modulation strength.     To obtain the spatial decay of the order parameter as a function of distance from the interface, we fit values of the ensemble average $\braket{\Delta_{mid}}$ for fixed disorder strength and different chain lengths \footnote{We find this method of fitting to be more accurate and unambiguous than directly fitting the curves in real space.}. We use histograms of $\Delta_{mid}$ for fixed system size and disorder strength to study the distributions of the induced OP. 
    
    Before moving on to inhomogeneous systems, it is useful to recall results for the periodic case, when the N chain hopping amplitudes are uniform, i.e. $t_i = -1$. The real space profile of the order parameter for the clean N-SC ring is shown in Fig.~\ref{fig:per-Delta}.  We find inverse distance behavior at zero temperature and exponential decay at finite temperatures (see Fig.~\ref{fig:per-scaling}). These results are in agreement with analytical calculations using Gor'kov's Green function method to compute $\Delta$ \cite{parks_proximity_2018, falk_superconductors_1963}.

\section{The proximity effect in disordered chains}\label{sec:Disordered}

\begin{figure}
    \includegraphics[width=0.45\textwidth]{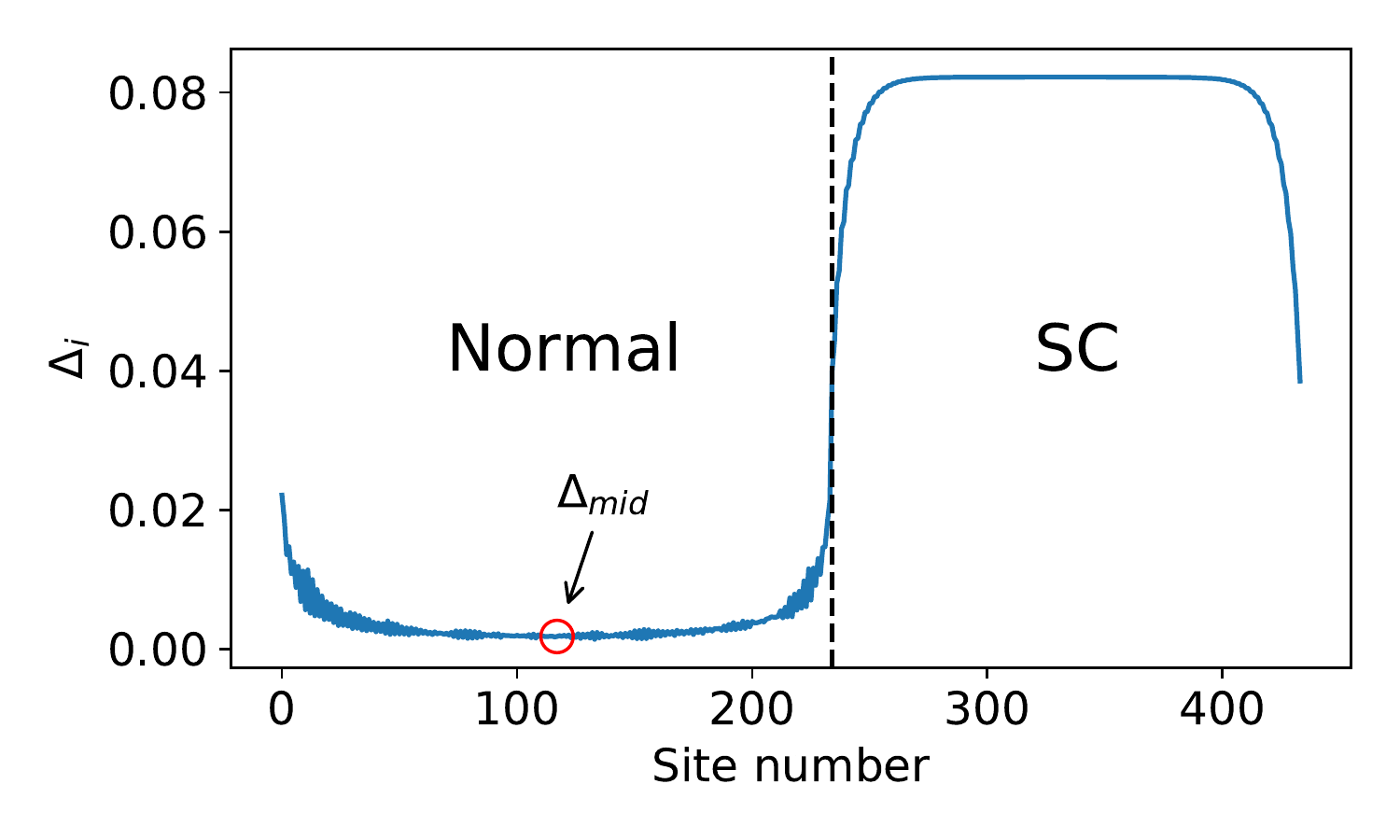}
    \caption{Superconducting proximity effect in a disordered one-dimensional metal: spatial profile of the superconducting order parameter $\Delta$ along the chain in a hybrid N-SC system with off-diagonal disorder in the normal region. Disorder strength: W/t = 0.08.}
    \label{fig:dis-Delta}
\end{figure}

As discussed, for example, by Pannetier and Courtois \cite{pannetier_andreev_2000}, in disordered non-interacting metals the proximity effect results from the Andreev reflections at the N-S interface combined with the presence of long range coherence of the metal. We now ask what happens in 1D systems, where the metallic state disappears upon the addition of disorder. Indeed, it is well known that adding arbitrarily small disorder in the one-dimensional periodic model leads to Anderson localization -- the Lyapunov exponent (inverse localization length) is non-zero for all eigenstates. In a finite chain, however, one can identify a weak disorder regime, in which the localization length $\xi(E)$ of single particle eigenstates is much larger than the system size $L$, i.e. $\xi \gg L$. Upon increasing the disorder strength one then has a crossover to a strong disorder regime, where the localization length is smaller than $L$, i.e. $ \xi < L$. For $T=0$, a third length scale, the inelastic (phase breaking) length scale, in this noninteracting model is infinite and thus plays no role.

\begin{figure*}
\centering
    \subfloat[\label{fig:weak-dis-scaling}]{
    \includegraphics[height=0.2\textwidth]{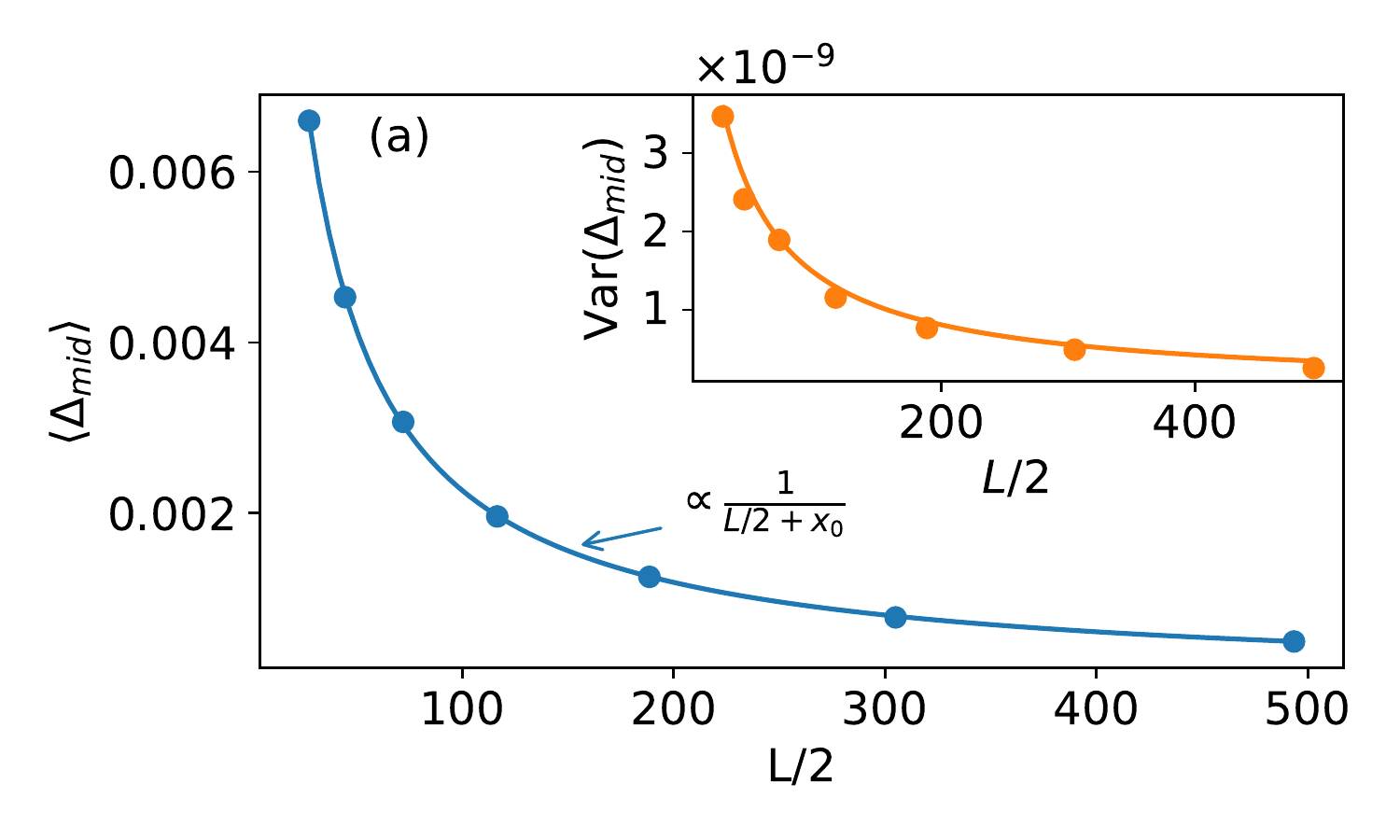}}
    \subfloat[\label{fig:weak-dis-hist}]{
    \includegraphics[height=0.2\textwidth]{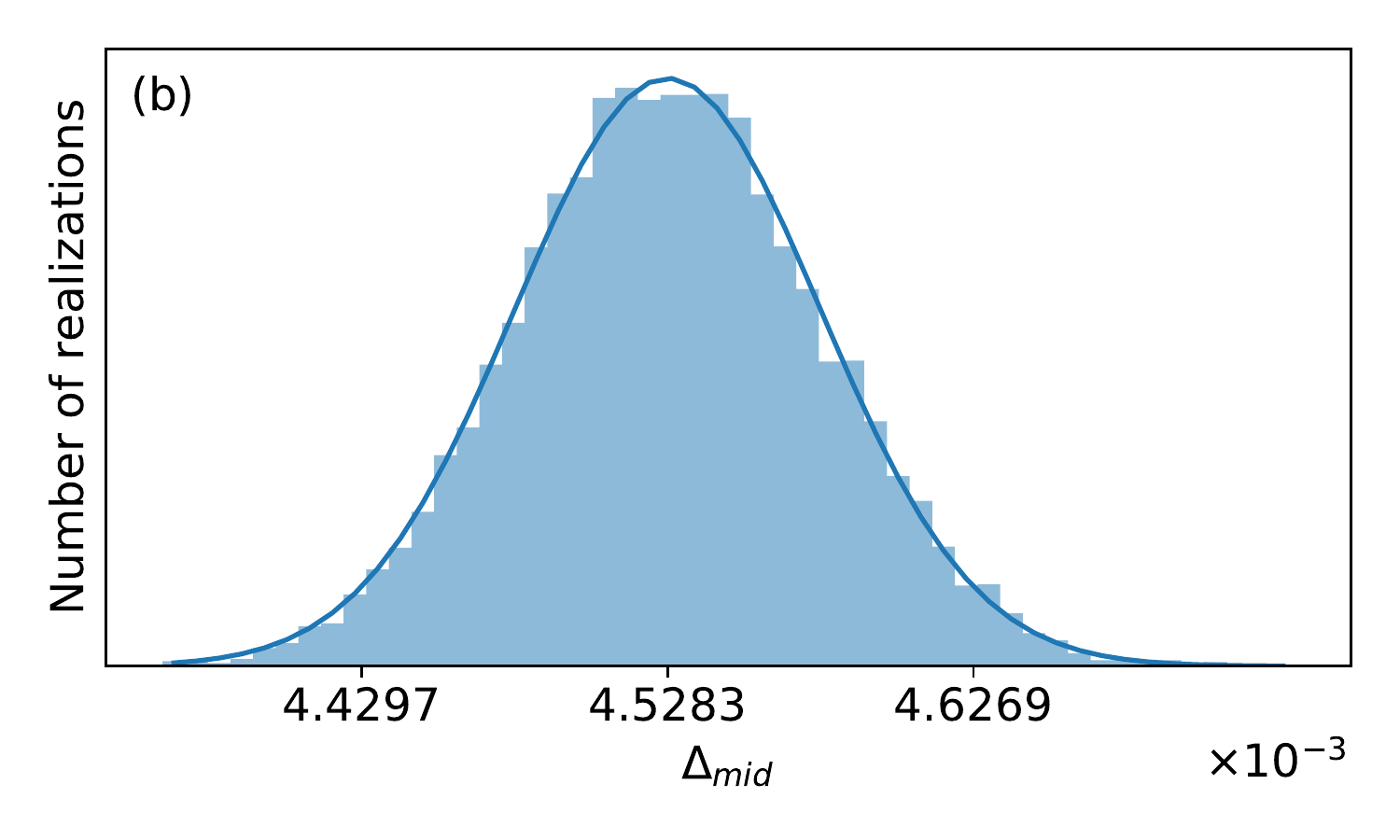}}
    \subfloat[\label{fig:weak-dis-w}]{
    \includegraphics[height=0.2\textwidth]{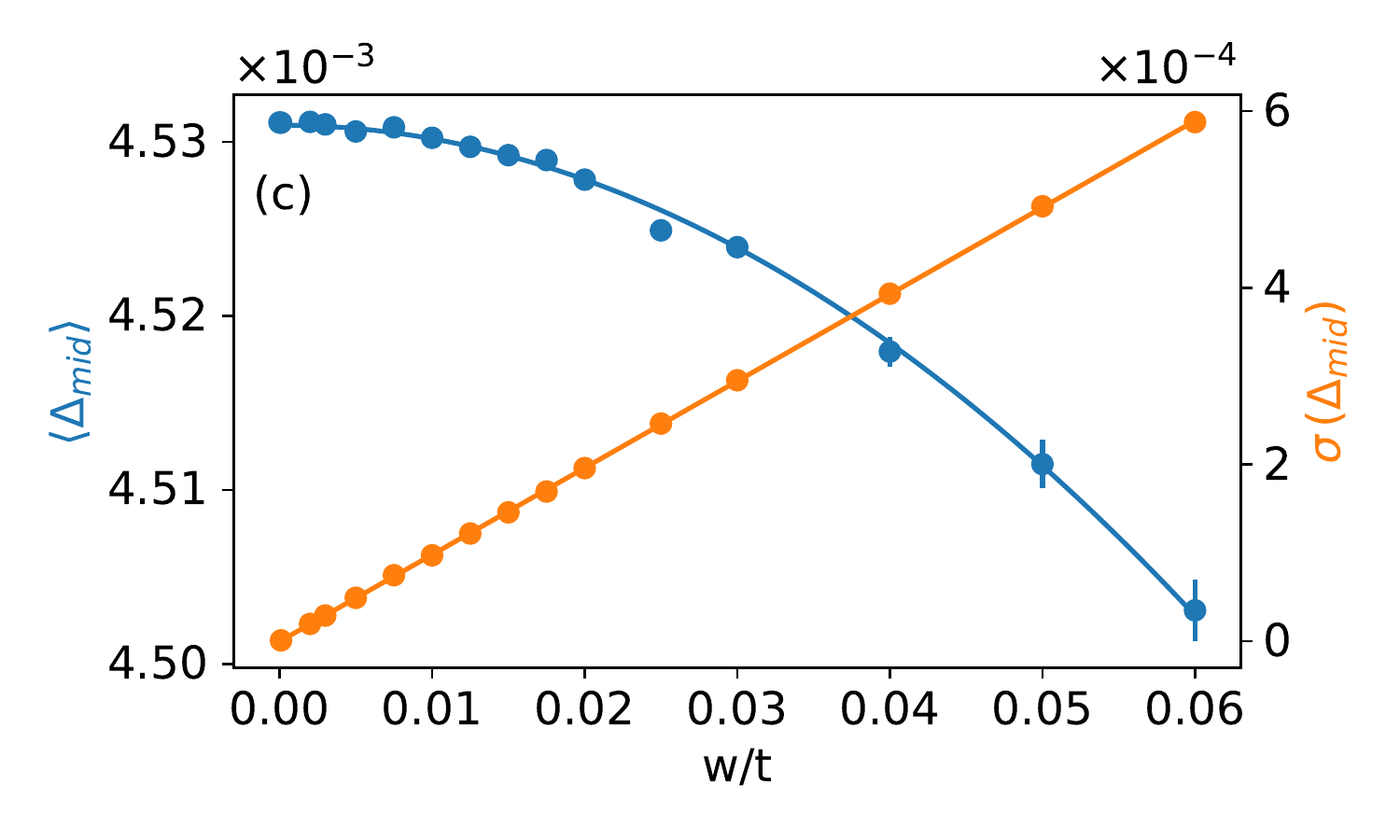}}
    
    \caption{Superconducting proximity effect in the weak disorder regime:
    (a) Finite-size scaling of $\braket{\Delta_{mid}}$ with the length of the normal region $L$ shows inverse law decay of $\Delta$ away from the interface in N--SC hybrid rings where N is weakly disordered. The variance of $\Delta_{mid}$ is also inversely proportional to the distance. Disorder strength: W/t = 0.005. Fit offset $x_0 = 10.4$. (b) Normal distribution of $\Delta_{mid}$ in N-SC rings in the weak-disordered regime. 10000 realizations, $W/t = 0.005$, $L = 90$, $L_{SC} = 200$. (c) The dependence of the mean and standard deviation of $\Delta_{mid}$ on the disorder strength $W/t$ in the weak-disordered regime.}
\end{figure*}

  In the following we present results corresponding to the two disorder regimes using the self-consistent theory outlined above. We show, firstly, that in the weak disorder case, the proximity induced superconducting order parameter (OP) decays as a power law of the distance from the interface. In contrast, the OP decays exponentially in the strong disorder regime. In addition, we obtain the full probability distributions of the OP and show how they differ in the two regimes.

We will consider the off-diagonal disordered Anderson model, in which the on-site terms are uniform and set equal to zero, while the hopping amplitudes $t_i=t+\eta_i$ are random. The independent random variables $\eta_i$ are  drawn from a uniform (box) distribution $P(\eta) = \frac{\Theta(\eta+\frac{W}{2}) - \Theta(\eta -\frac{W}{2})}{W}$ where $W/t<2$ is the disorder strength, and $\Theta(\eta)$ is the Heaviside step function. The width of the distribution is restricted to be less than 2, so that the hopping amplitudes $t_i$ are all strictly negative. Although we work with a box distribution, the exact form of the randomness is not expected to matter for our results, which should also hold for more general random distributions \footnote{In preliminary calculations on the diagonal Anderson model, we found the same qualitative behaviour}.

\subsection{Weak Disorder Regime}
In a weakly disordered finite chain, one can compute the (sample-dependent) corrections to the energies and wavefunctions of the clean system using perturbation theory. In this limit, the semi-classical viewpoint -- in which the principal effect of the randomness is to randomize the phase of the wavefunctions -- is useful. That wavefunctions in reality remain extended, can be readily seen from the fact that the average $probability$ at each site $\langle \psi^2_n(i)\rangle$ tracks the values obtained in the clean system. We therefore use the term quasi-extended to denote this type of wave function. Fig. 2 shows a typical order parameter profile in the weak disorder regime. Fig. 3 shows our main numerical results. The order parameter decays as a power law away from the interface into the normal part of the ring (Fig. \ref{fig:weak-dis-scaling}). $\Delta_{mid}$ for a given system size and disorder strength is normally distributed (Fig. \ref{fig:weak-dis-hist}). $\Delta_{mid}$ on average differs from the clean case by a term proportional to $W^2$ (Fig. \ref{fig:weak-dis-w}).

These observations can be explained by means of perturbation theory in the variables $\eta_i$. In the clean limit, the eigenfunctions and energies of the hybrid chain system have been analytically studied in \cite{buttiker_flux_1986, cayssol_isolated_2003}. Considering a BdG type model in which they fixed the OP in the superconductor to a constant value, these authors showed that there is a finite, constant, density of states within the gap.  These states should exist even after relaxing the constraint on the OP, as we do in our self-consistent approach. These eigenstates are of interest in the following perturbative argument for the induced OP within the N chain. 
We write the solutions of Eq.~\eqref{eq:BdG} in terms of the coefficients for the clean system $u_n$, $v_n$ and corrections to these, $ \delta u_n$ and $\delta v_n$. Within perturbation theory, the correction terms $\delta u_n$ and $\delta v_n$ up to second order in $\eta_i$ are kept. At zero temperature, we have an expression for the order parameter at a given site (we suppress the index $i$) from Eq.~\eqref{eq:self-con} \footnote{We have exploited the fact the $u$, $v$, and $\Delta$ can be chosen to be real in the absence of magnetic fields and that all $\epsilon_n>0$},
\begin{align}
    \Delta &= \sum_n (v_n + \delta v_n)(u_n +\delta u_n) \nonumber \\
    &= \Delta_{0} + \sum_n v_n \delta u_n + \delta v_n u_n + \delta v_n \delta u_n ,
\end{align}
where $\Delta_0$ is the OP in the clean case. The normalization condition $\sum_n |u_n|^2 + |v_n|^2 = 1$ leads to
\begin{align}
    \sum_n u_n \delta u_n + v_n \delta v_n = -\frac{1}{2}\sum_n(|\delta u_n|^2 + |\delta v_n|^2).
\end{align}
Taken together,
\begin{align}
    \Delta - \Delta_{0} = \sum_n \bigg\{& (v_n-u_n)(\delta u_n - \delta v_n)\nonumber\\ &- \frac{1}{2}(|\delta u_n|^2 + |\delta v_n|^2)\bigg\}  .
        \label{eq.deltaweak}
 \end{align}
 Averaging over the disorder then yields
\begin{align}
    \langle \Delta-\Delta_{0} \rangle \approx  - \frac{1}{2}\sum_n \langle(|\delta u_n|^2 + |\delta v_n|^2), \rangle
    \label{eq.deltaweak2}
\end{align}
  where we neglected the contribution of the first term in \eqref{eq.deltaweak} compared to that of the second term. This follows because the averages of $\delta u_n$ and $\delta v_n$ are very small (the linear corrections in $\eta$ average to zero, and the second order corrections are small)  compared to the average of $|\delta u_n|^2$ and $|\delta v_n|^2$. Note that the correction term due to disorder in \eqref{eq.deltaweak2} is always negative and proportional to $W^2$.

For a chain of length $L$, Eq.~\eqref{eq.deltaweak} is a sum of $L/2$ random variables of variance proportional to $W^2$. Therefore, we can  invoke the central limit theorem to see that the distribution of $\Delta_{mid}$ must be a Gaussian, with a width proportional to $W$, centered at $\Delta_{0} - c (W/t)^2$ where $c$ is a constant. The scaling of the width of the distribution with system size is given by the product of a factor of $1/L$ (normalization), and a factor $\sqrt{L}$ (from the variance of the sum over $L$ random variables). These features are observed in the inset of Fig. \ref{fig:weak-dis-scaling} and in Fig. \ref{fig:weak-dis-w}.


We now discuss the spatial decay of the OP, which we compute from the dependence of the average OP value at the midpoint of the chain, $\braket{\Delta_{mid}}$, for different lengths $L$. As shown in Fig.~\ref{fig:weak-dis-scaling}, $\braket{\Delta_{mid}}$ has an inverse distance or $1/L$ dependence. This is expected in view of the weak localization physics we expect in this regime. According to theory, the averaged density-density correlations are known to obey a diffusion equation \cite{akkermans_mesoscopic_2007}. These correlations therefore decay as the inverse power of distance, similar as in a pure metal. In our present context, this property implies that the spatial dependence of the proximity induced averaged pair correlation function will be similar to that of the clean system. In other words, it should fall off with the inverse of the distance from the interface, $\Delta(x) \sim 1/x$ \cite{parks_proximity_2018}. These $T=0$ properties should carry over at finite temperatures as long as the phase breaking length scale remains large compared to the system size.

\begin{figure*}
    \centering
    \subfloat[\label{fig:strong-dis-scaling}]{
    \includegraphics[height=0.2\textwidth]{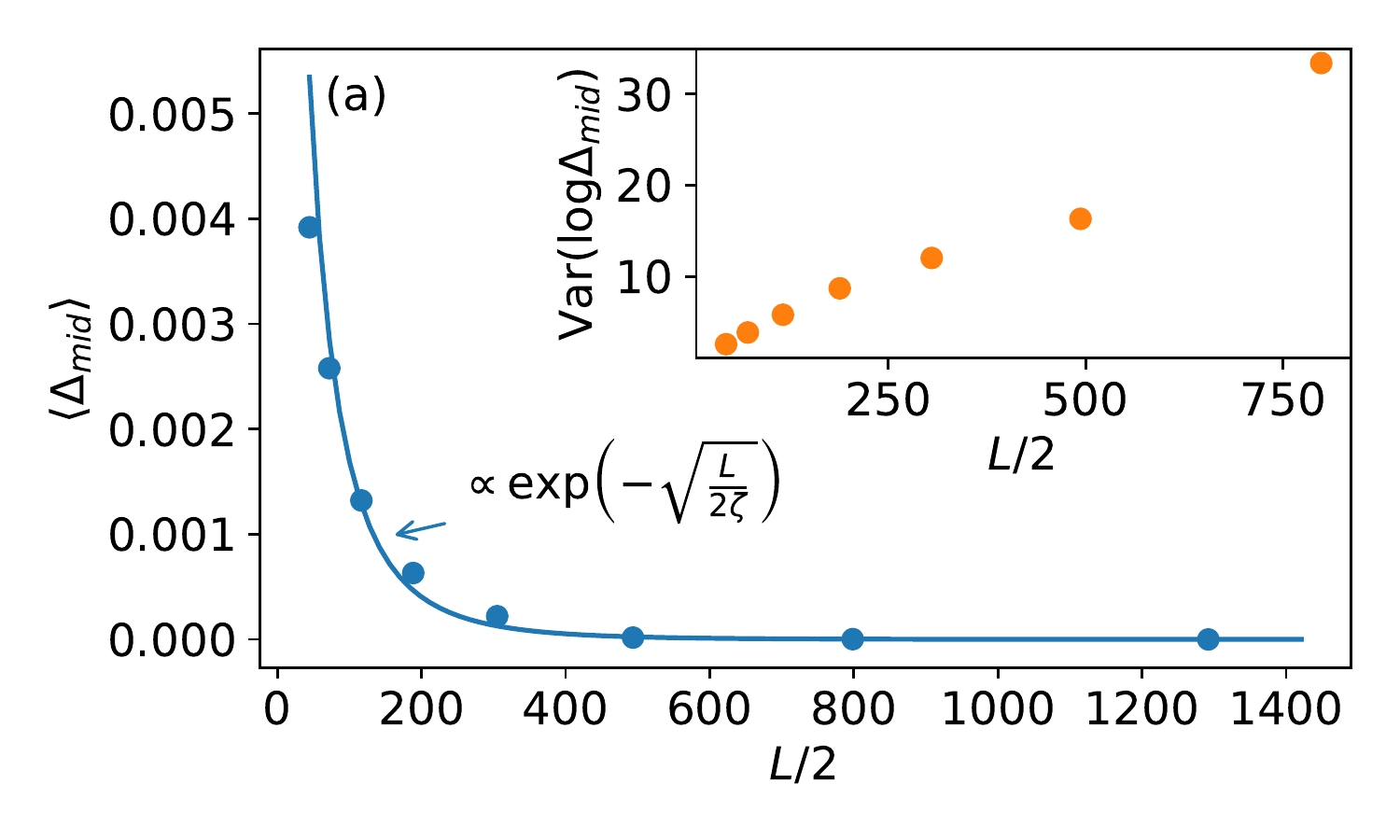}}
    \subfloat[\label{fig:strong-dis-hist}]{
    \includegraphics[height=0.2\textwidth]{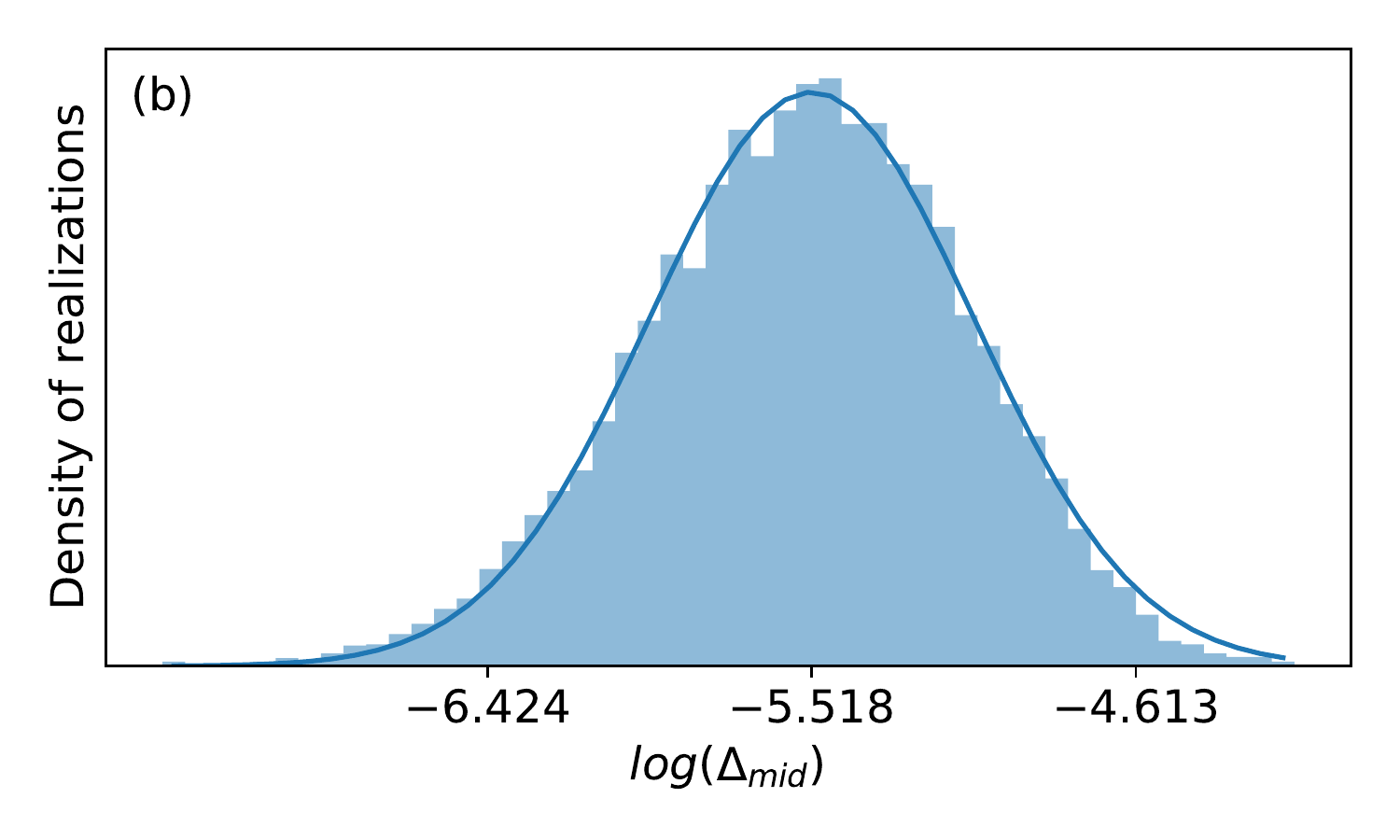}}
    \subfloat[\label{fig:strong-dis-w}]{
    \includegraphics[height=0.2\textwidth]{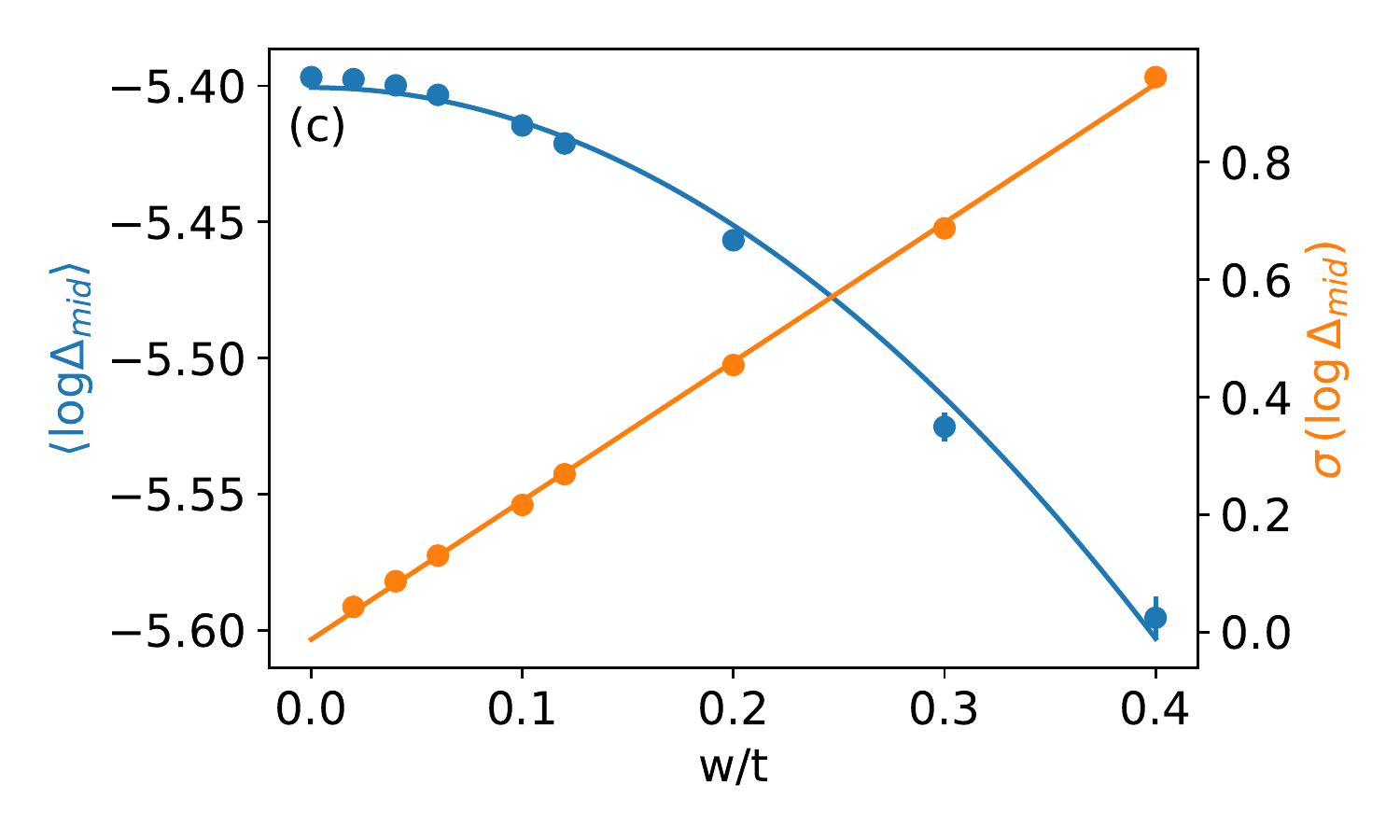}}
    
    \caption{Superconducting proximity effect in the strong disorder regime:
    (a) Finite-size scaling of $\braket{\Delta_{mid}}$ with the length of the normal region $L$ shows stretched exponential decay of $\Delta$ away from the interface in N--SC hybrid rings where N is strongly disordered. Disorder strength: W/t = 0.7.  $\zeta = 8.4$ (b) Log-normal distribution of $\Delta_{mid}$ in N-SC rings in the strong-disordered regime. 9000 realizations, W/t = 0.2, $L = 90$, $L_{SC} = 200$. (c) The dependence of the mean and standard deviation of $\log(\Delta_{mid})$ on the disorder strength $W/t$ in the strong-disordered regime.}
\end{figure*}

\subsection{Strong Disorder}

The strong disorder regime corresponds, in our model, to values of $W/t$ of order 0.1 or larger. In this regime, the wave functions have an exponentially decaying envelope function. The proximity effect is expected to be short-ranged, in contrast to the weak disorder regime. Indeed this is observed in Fig.~\ref{fig:strong-dis-scaling}, where the value of $\langle\Delta_{mid}\rangle$ for a fixed disorder strength $W/t=0.7$ is plotted as a function of system size $L$. The characteristic decay length decreases as $W$ is increased, in accordance with the theoretically predicted exponential behavior.

Fig.~\ref{fig:strong-dis-hist} shows the distribution function of $\langle\Delta_{mid}\rangle$ in the strong disorder regime. This distribution is well-described by a log-normal form, i.e. the variable $y=\log(\Delta_{mid})$ is distributed according to a Gaussian,
\begin{eqnarray}
P(y) = C e^{-(y-y_0)^2/2\sigma(W,L)},
\end{eqnarray}
where $C$ is a normalization constant. The width of the Gaussian, $\sigma(W,L)$, is found to grow linearly with $W$ and with the system size $L$, as shown in Fig.~\ref{fig:strong-dis-w}. 

We now present an argument for these observations.  Due to the bipartite character of the random hopping Hamiltonian, the single particle states exactly at $E=0$ are known to have a stretched exponential form \cite{theodorou_extended_1976,economou_static_1981}, meaning a spatial decay faster than a power law but slower than a pure exponential. This is most readily shown by using the tight-binding equations for $E=0$ to relate the wave function amplitude $\psi(i)$ in the interior of the chain to the wave function on the boundary $\psi(0)$ and $\psi(1)$. For a site located at a distance $2m$ from the boundary, the local wave function amplitude is
\begin{eqnarray}
\psi_{E=0}(2m) \propto \prod_{1\leq l \leq m} (-1)^m \frac{t_{2l}}{t_{2l-1}}
\end{eqnarray}
A similar relation holds for the state on the odd sublattice. From the above it is clear that the logarithm of the $E=0$ wave function at the midpoint of the chain of length $L=2M$ can be written as a sum, 
\begin{eqnarray}
\log\psi_{E=0} = \sum_{i=1}^M x_i + const,
\label{eq.strongdis}
\end{eqnarray}
where the random numbers $x_i$ are related to the hopping amplitudes by $x_i=\log(t_{2i}/t_{2i-1})$ \cite{soukoulis_off-diagonal_1981}. The above equation shows that, according to the central limit theorem, $\ln\psi$ is a Gaussian distributed random variable. Its variance increases with the number of $x_i$, that is, is proportional to the chain length $L$.

In the strong disorder regime, we assume that  $\Delta_{mid}$ can be approximately written as
\begin{eqnarray}
\Delta_{mid} \sim u_{L/2}^{E = 0} v^{E = 0}_{L/2} 
\label{eq:deltaapprox}
\end{eqnarray}
i.e. that the contributions from finite energy states in the sum \eqref{eq:self-con} can be neglected, so the OP at the midpoint is determined principally by the wave functions $u$ and $v$ at $E=0$.
This simplification can be justified as a result of two factors:  firstly the fact that there is a singularity (in finite systems, a peak) in the density of states at this energy \cite{khan_probing_2020}. Secondly, the localization length is largest at the band center  and decreases for energies away from $E=0$ \cite{pichard_one-dimensional_1986, izrailev_anomalous_2012}, so that contributions due to finite energy states should be small. 
The OP at the midpoint is thus determined by the wave functions $u$ and $v$ at the midpoint of the chain which both have the form given by Eq.\ref{eq.strongdis}, differing only in the values of the prefactor. The central limit theorem applied to the logarithm of the product $uv$ tells us that $\log(\Delta_{mid}) \sim \log(\psi)$ must have a Gaussian distribution of width proportional to $W$,  increasing with system size as $\sqrt{L}$. This is in agreement with the results shown in Fig.~4. Note that for strong disorder the distribution width of the OP $grows$ with the system size, in contrast with the distribution in the weak disorder limit where it $decreases$ with $L$. The proximity induced OP is clearly a strongly fluctuating quantity, analogous to the distribution of values of the resistivity in 1D systems \cite{abrahams_resistance_1980,soukoulis_off-diagonal_1981}. 

To conclude this section, we have shown that extended states lead to a power law decay of the OP and a Gaussian distribution in the weak disorder regime, whereas localized states lead to a stretched exponential decay and a log-normal distribution in the strong disorder regime. We have checked that these features in the induced order parameter in off-diagonal Anderson model carry over to the the diagonal Anderson model except that for high values of $W$, the OP decay is fit better by an exponential rather than a stretched exponential.

\section{Proximity Effect in the Fibonacci chain}\label{sec:Fibonacci}

\begin{figure}
    \centering
    \subfloat[    \label{fig:FC-Delta}]{
    \includegraphics[width = \columnwidth]{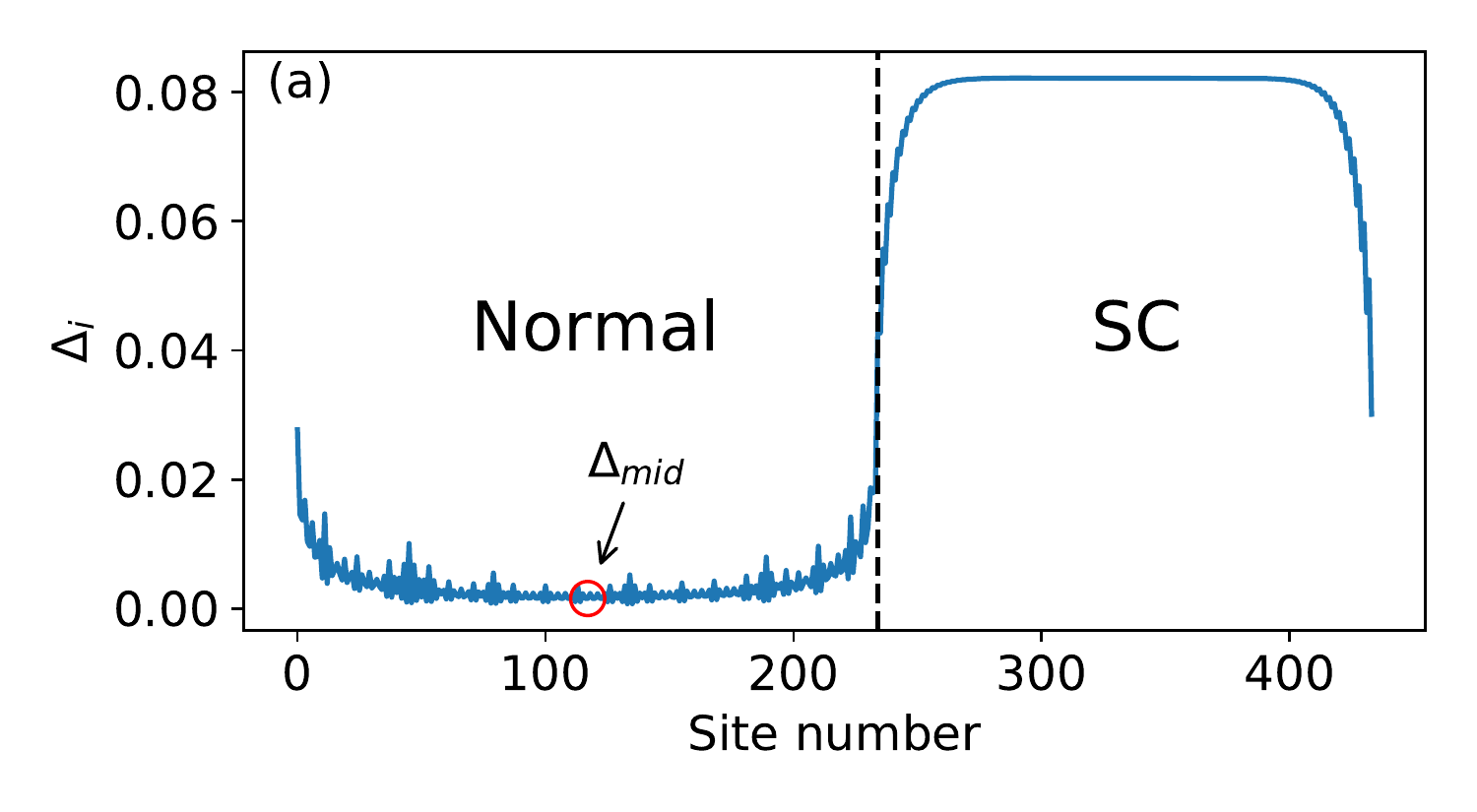}}\\
    \subfloat[    \label{fig:fractal}]{
    \includegraphics[width = \columnwidth]{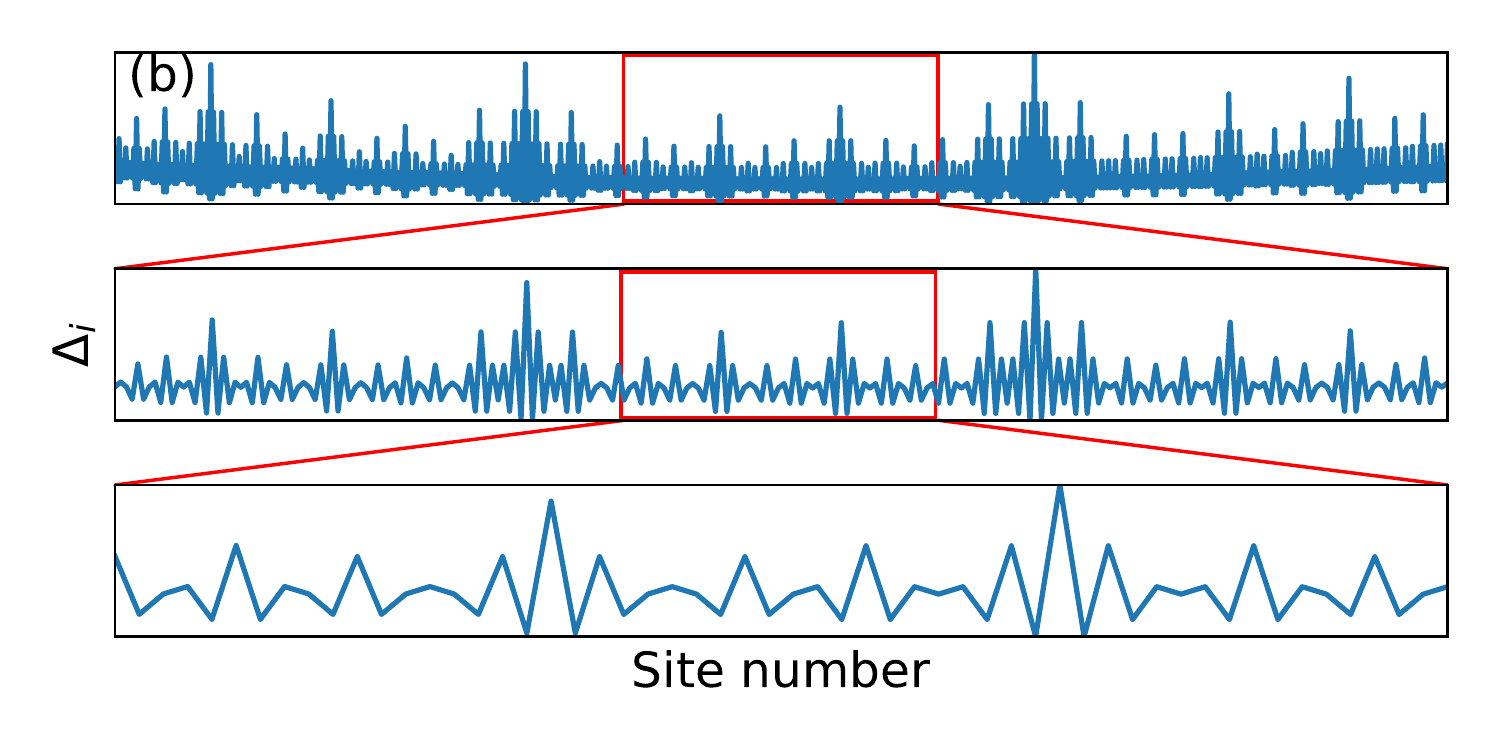}}\\
     \subfloat[    \label{fig:FC-sketch}]{
    \includegraphics[width = \columnwidth, trim = -15 0 0 0, clip]{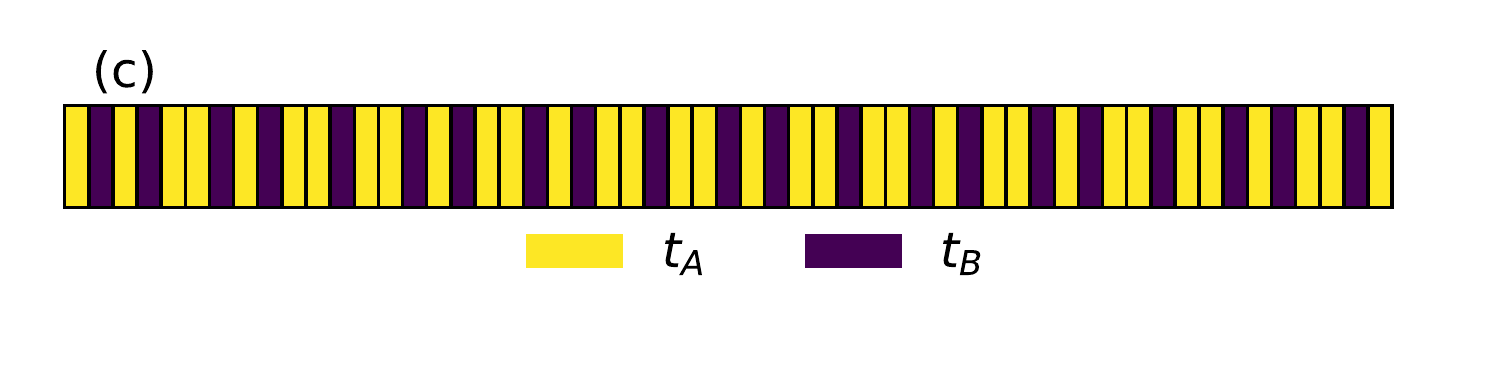}}
    \caption{Superconducting proximity effect in a Fibonacci chain: (a) Real-space profile of the superconducting order parameter when the normal segment is a Fibonacci chain. Modulation strength: $W/t = 0.1$ (b) Superconducting order parameter profile in the central 987 (top), 233 (middle), and 55 (bottom) segments of the Fibonacci chain in a hybrid Fibonacci-SC ring. The order-parameter profiles are self-similar upon scaling by the renormalization parameter $\tau^3$. The data in panel (b) is taken from a hybrid ring with 2585 and 200 sites in the normal and superconducting regions respectively. (c) The hopping sequence corresponding to the sites shown in the last panel in (b).}
\end{figure}

We now consider a quintessential example of a one-dimensional quasiperiodic system---the Fibonacci chain. The Fibonacci chain is the most influential and best studied example of a 1D quasicrystal. It exhibits many of the most interesting novel features of quasicrystals including a singular continuous spectrum\cite{ostlund_one-dimensional_1983, kohmoto_localization_1983, hiramoto1992electronic}, topological edge states \cite{rai_proximity_2019}, critical states \cite{mace_critical_2017}, and discrete scaling symmetries \cite{mace_fractal_2016}. We consider the off-diagonal model, in which the hopping amplitudes $t_i$ take one of two values, $t_A$ and $t_B$. We start by considering chains which are obtained by repeated application of the map $\sigma:$ $\sigma(t_A) \to t_A\,t_B, \sigma(t_B)\to t_A$ to the initial sequence $\{t_B\}$. To illustrate this, the first few applications yield $\{t_B\}\to\{t_A\}\to\{t_A\,t_B\}\to\{t_A\,t_B\,t_A\}$, and so on. These finite sequences obtained after $n$ applications of $\sigma$ are called approximants of the Fibonacci quasicrystal. The number of hoppings in an approximant is a Fibonacci number $F_n = 1, 2, 3, 5, 8 \ldots$. In an approximant of length $F_n$, the ratio of the number of $A$-bonds to the number of $B$-bonds is $\tau_n = F_{n-1}/F_{n-2}$ and tends to the golden mean, $\tau=\frac{1+\sqrt{5}}{2} \approx 1.618$ as the chain length tends to infinity. We will see below, when we come to the study of fluctuations that that these are not the only allowed Fibonacci approximant sequences---there are, in fact, $F_n$ different approximants corresponding to a given generation $n$.

We will consider henceforth the case where $t_A \leq t_B$.  The case $t_A=t_B$ corresponds to the periodic chain, and the difference $W = t_B - t_A$ gives a measure of the strength of the quasiperiodic modulations. For a given value of $W$, we choose $t_B$ and $t_A$ such that the average of the hoppings over the chain is unity $\implies \frac{1}{F_n}(F_{n-1} t_A + F_{n-2} t_B) = -1$. The spectrum and wave functions of this hopping Hamiltonian have been extensively studied, and it is well-known that all states are multi-fractal and critical \cite{hiramoto1992electronic}. Such states are characterized by very large fluctuations, and wave function amplitudes decay with different power laws, depending on the local environment. This has been checked by numerical calculations and by explicit computations within a perturbative renormalization group approach \cite{mace_fractal_2016}. The state for $E=0$ has been studied in detail in \cite{mace_critical_2017} and is of particular interest for the proximity effect, as discussed next.

\subsection{OP profile in Fibonacci approximants}

\begin{figure}
    \centering
    \includegraphics[width = \columnwidth]{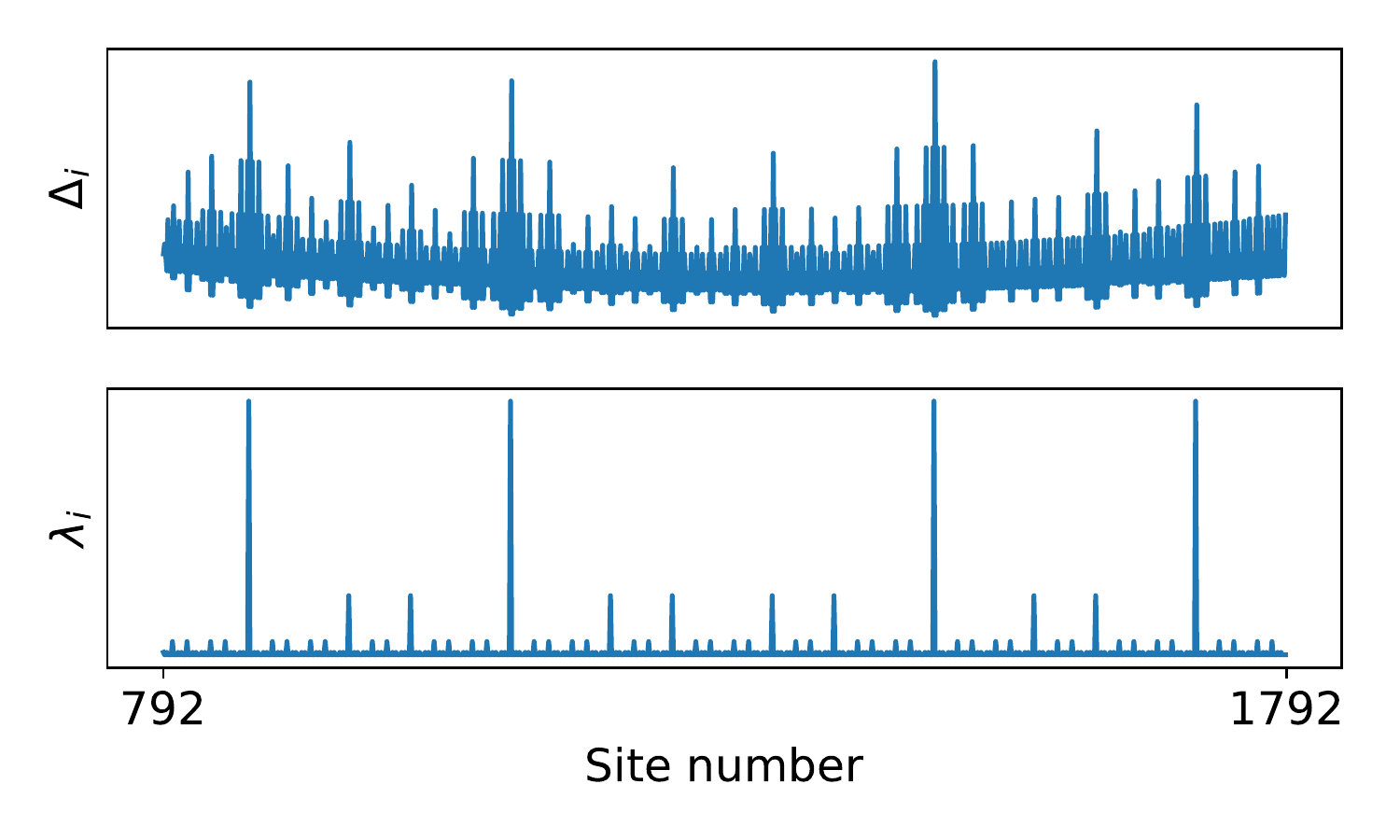}
    \caption{The local order parameter and its correlation with the symmetry of the local environment: the peaks in $\Delta_i$ occur at sites with high reflection symmetry represented by $\lambda_i$.}
    \label{fig:resonators}
\end{figure}

Fig.~\ref{fig:FC-Delta} shows the spatial behavior of the superconducting order parameter in a hybrid ring composed of a Fibonacci chain coupled to a BCS superconductor. This order parameter profile displays a self-similarity, which can be seen in Fig.~\ref{fig:fractal} where we zoom into successively smaller sections in the center of the Fibonacci chain. The plots show a central region where the number of sites is reduced successively by a factor $\tau^3\;$ \footnote{The $\tau^3$ scaling factor is a general feature of the Fibonacci chain as reported in \cite{chakrabarti_exact_1989}}.  

Local environment plays an important role in the value of the OP on a given site. The lowest panel of Fig.~\ref{fig:fractal} shows the hopping sequence for the region shown in Fig.~\ref{fig:FC-sketch}, with $t_A$ and $t_B$ shown by yellow (resp. black) bands. One sees a correlation between the heights of OP peaks and the local environment, described as follows. Let us define $\lambda_i$ as the distance up to which reflection symmetry is present around a given site $i$: specifically $\lambda_i$ is the smallest whole number $d$ such that $t_{i+1+d} \neq t_{i-d}$. In of Fig. \ref{fig:resonators}, notice that peaks in $\lambda_i$ coincide with higher values of $\Delta_i$. This type of characterization of local environments was presented in \cite{rontgen_local_2019} where they find that edge states with certain energies localize in such high symmetry regions, which they call \emph{local resonators}.

This self-similarity and the local environment dependence of the OP have simple explanations in terms of the theoretical description of Fibonacci chain eigenstates in \cite{mace_fractal_2016} and \cite{mace_critical_2017}. We assume, as in \eqref{eq:deltaapprox}, that the sum over states can be replaced by the $E=0$ contribution. The OP is thus determined by the structure of the $E=0$ wavefunctions which is well-known. In the limit of strong quasiperiodic modulation, these wavefunctions tend to have their support concentrated on the sites which are surrounded on either side by $t_A$ bonds (and were called ``atom" sites in the renormalization group (RG) approach due to Niu and Nori and Kalugin et al \cite{niu_renormalization-group_1986, kalugin_electron_1986}). Under a renormalization transformation, the absolute value of the $E=0$ wave function of a site in the $n$th generation chain is related to that of a site in the $(n-3)$th generation by the recursion formula
\begin{eqnarray}
\vert \psi_{E=0}^{n}(i) \vert = \sqrt{\overline{\lambda}} ~\ \vert\psi_{E=0}^{n-3}(i')\vert
\end{eqnarray}
where $i$ and $i'$ are the site indices of the old and the new (renormalized) chain, and $\overline{\lambda}$ is a wavefunction rescaling factor which can be computed as a function of the hopping parameters \cite{mace_fractal_2016}. The superconducting OP is given by the product of two such wavefunctions. The highest amplitude is found for the sites which remain after the largest number of RG transformations. Under RG transformation, the number of such sites is reduced by a factor $\tau^3$.
 The distance out to which the site possesses reflection symmetry increases by the same factor. Thus the RG theory of the Fibonacci chain explains the numerical observations of a) self similarity of the OP, and b) the correspondence between the OP and the local environment.

\begin{figure*}
    \centering
    \subfloat[\label{fig:FC-scaling}]{
    \includegraphics[height=0.2\textwidth]{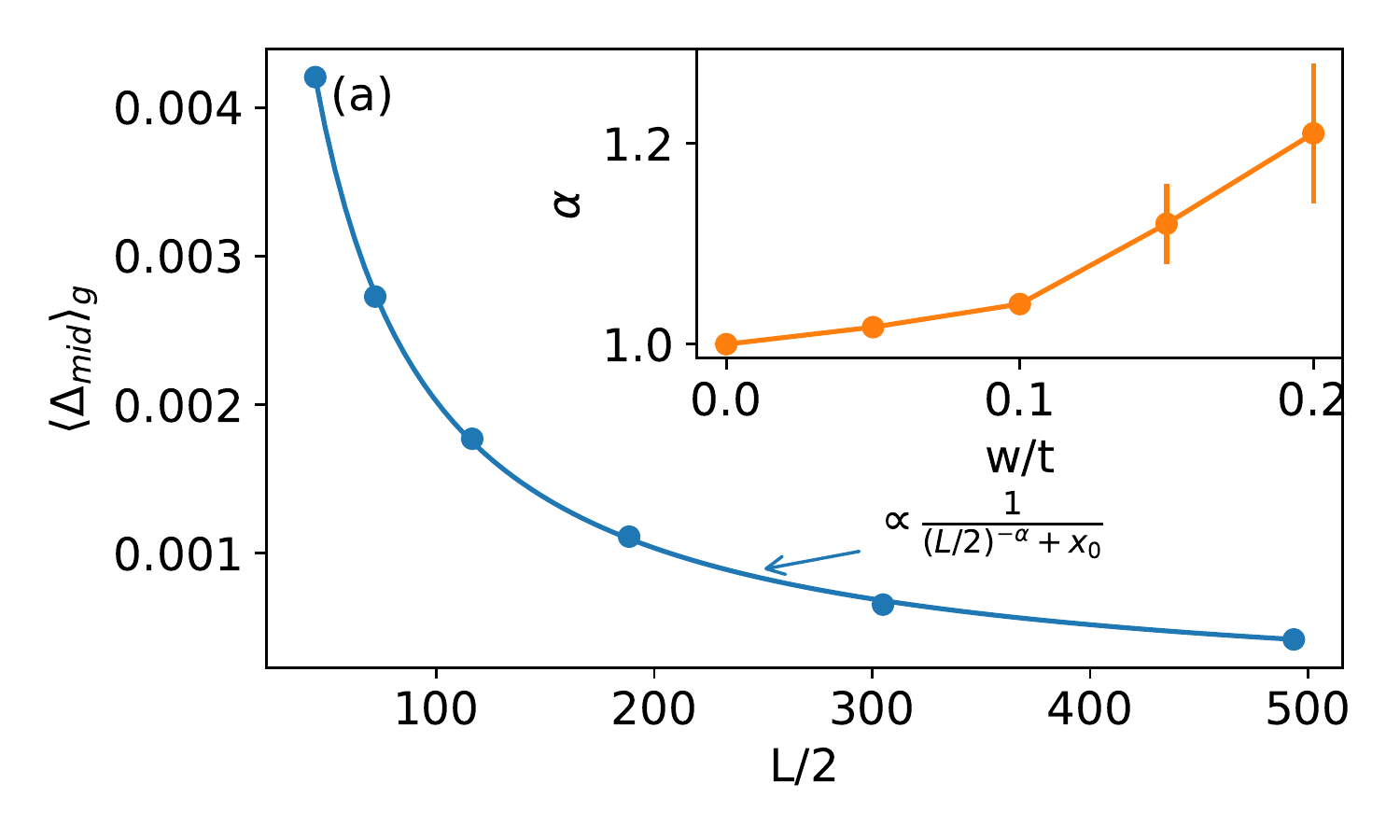}}
    \subfloat[\label{fig:FC-hist}]{
    \includegraphics[height=0.2\textwidth]{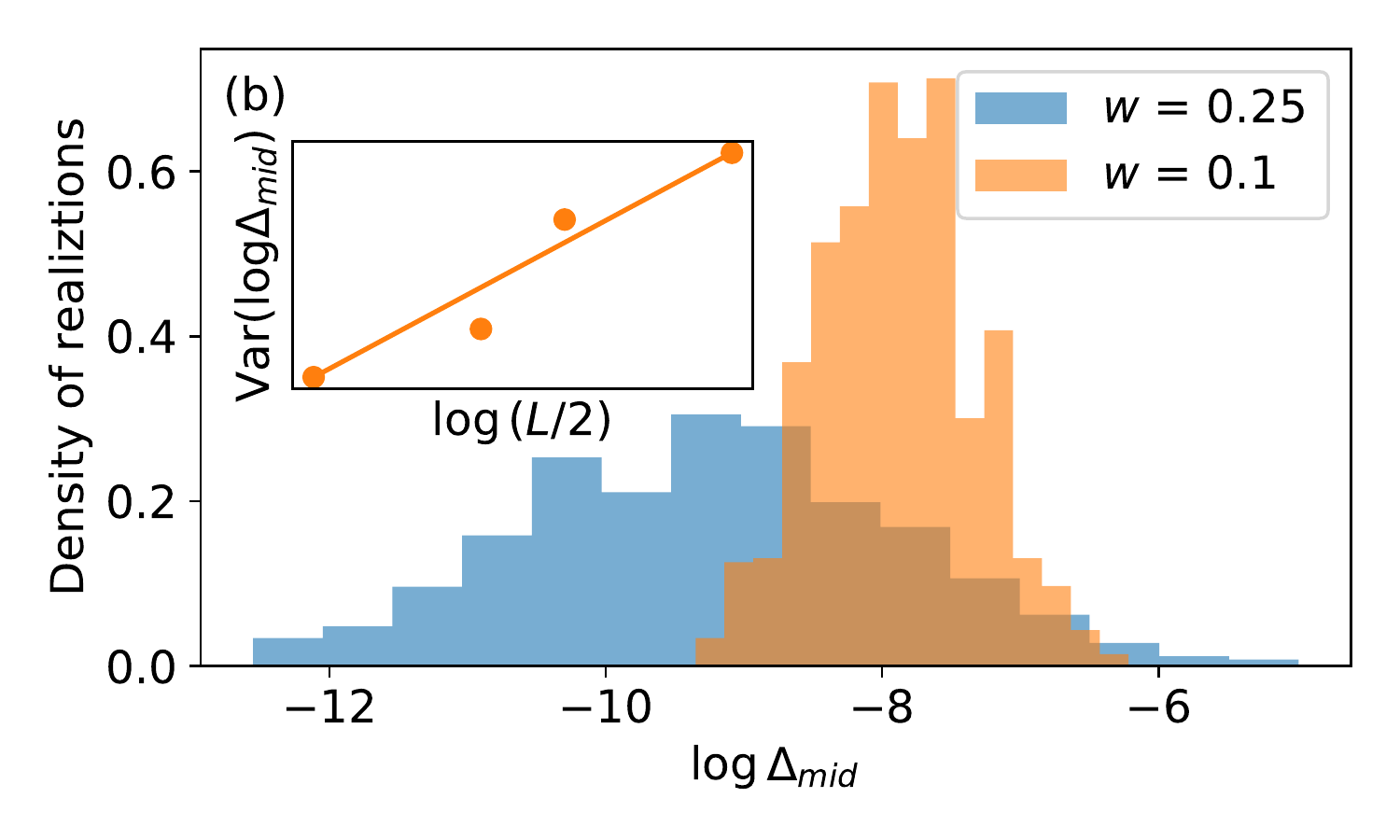}}
    \subfloat[\label{fig:FC-w}]{
    \includegraphics[height=0.2\textwidth]{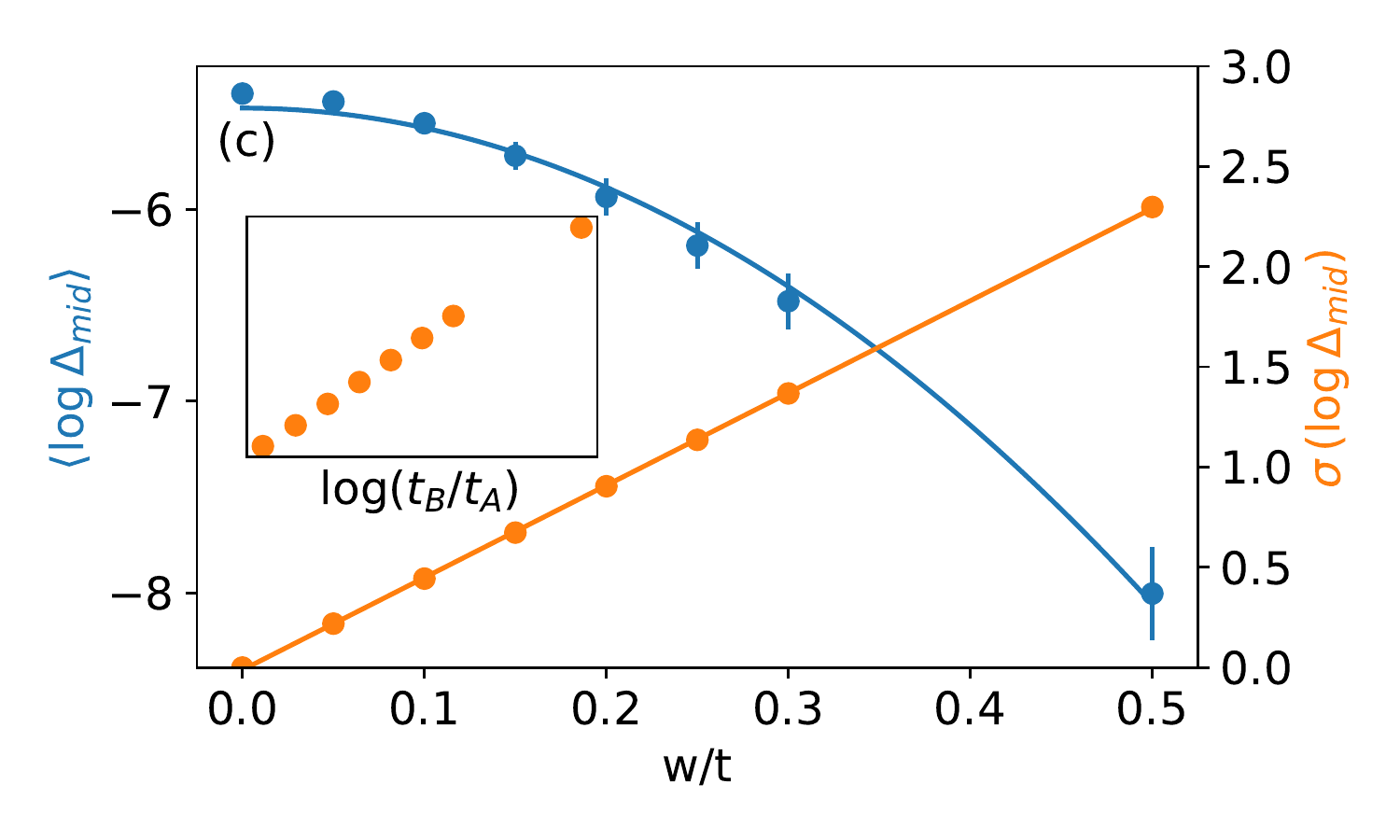}}
    
    \caption{Superconducting proximity effect in a Fibonacci chain:
    (a) Finite-size scaling of $\braket{\Delta_{mid}}$ with the length of the normal region $L$ shows power law decay of $\Delta$ away from the interface in N--SC hybrid rings where N is a Fibonacci chain. Modulation strength: W/t = 0.005. Fit offset $x_0 = 10.85$ (b) The distribution of $\log\Delta_{mid}$ in N-SC rings is symmetric when N is a Fibonacci chain. $L = 988$, $L_{SC} = 200$. (c) The dependence of the mean and standard deviation of $\Delta_{mid}$ on the modulation strength $W/t$. Inset: The functional dependence of the width, $\sigma(\log \Delta_{mid})$, on the ratio of the hopping parameters.}
\end{figure*}

\subsection{Fluctuations of the OP}

The chains built by substitution of the previous section are only one of a family of approximants: there are exactly $F_n$ Fibonacci approximants of length $F_n$. The local order parameters in the Fibonacci chain fluctuate according to the structure of the chain considered.
 A systematic way to generate all approximants of length $F_n$ uses the characteristic function,
\begin{align}
\chi_j = sign\left(\cos(2\pi j \tau_n^{-1} + \phi) - \cos(\pi \tau_n^{-1})\right).
\end{align}
$\chi_j$ gives the $j$th hopping, where we identify $-(+)$ with $t_A(t_B)$, and $j = 1, 2, \ldots F_n$. By varying $\phi$ throughout the interval $[0,2\pi)$, we recover all approximants of the given length. The difference between chains generated by changing $\phi$ is small and consists of \emph{phason flips}---exchanging a pair of bonds $t_B t_A \to t_A t_B$. 
The variation of the induced OP as a function of $\phi$ was studied in \cite{rai_proximity_2019}. There it was shown that varying $\phi$ leads to complex oscillations of the OP, with periods given by the topological indices of the gaps of the Fibonacci chain spectrum. 

One can plot the distribution function of the OP, just as we have done for the randomly disordered case. However, there is a significant difference between the models: whereas in the disordered system one can generate as many realizations of the chain as one wishes, for the Fibonacci chain there are only $F_n$ realizations of a chain of length $F_n$. It is therefore necessary to go to very large systems to obtain a good fit for the decay law of $\langle\Delta_{mid}\rangle$ and to fit the distribution function to a smooth form. In this work, the biggest system that we have studied consists of $N=1,597$ bonds. 

\subsubsection{Decay law of the typical value of the order parameter} In this subsection we consider the properties of the typical value of the OP $\langle\Delta_{mid}\rangle_g$---after averaging over all values of $\phi$ \footnote{ $\braket{\bullet}$ refers to the geometric mean of $\bullet$}---and focus in particular on its spatial decay away from the N-SC interface. 
 To study the spatial decay of $\langle\Delta_{mid}\rangle_g$ as one moves away from the interface, we compute this quantity for chains of different lengths $L$. Plotting it as a function of length $L/2$, as shown in Fig.~\ref{fig:FC-scaling} yields a power law with $\langle\Delta_{mid}(n)\rangle_g \sim n^{-\alpha}$, where the exponent $\alpha$ depends on the modulation of hopping amplitudes. 
 
The observed power law decay is consistent with the presence of critical states: the effective exponent is non-universal and depends on the average values of the density of states and the states close to the Fermi level. An explicit calculation is outside the scope of the present discussion. We remark simply that when the ratio $t_A \rightarrow t_B$, the power $\alpha\rightarrow 1$, i.e. approaching the decay law for the simple non-modulated chain (see inset of Fig.~\ref{fig:FC-scaling}). As one might expect, increasing the strength $W$ of the quasiperiodic modulation results in wave functions becoming less extended, leading to a faster spatial decay of the OP as one moves away from the interface. 

\subsubsection{Distribution of the OP}
The Distribution of the induced order parameters are shown in Fig.~\ref{fig:FC-hist} for a chain of length $L=988$ for two different values of the modulation strength $W$, while the dependence of its mean and standard deviation on the disorder strength is shown in Fig.~\ref{fig:FC-w}. More precisely, Fig.~\ref{fig:FC-hist} shows distributions of the logarithm of $\Delta_{mid}$, which are more symmetric. This can be explained by an argument, which suggests that $\Delta_{mid}$ should tend to a log-normal distribution. We approximate the sum in Eq.~\eqref{eq:self-con} for $\Delta_{mid}$ by keeping only the $E=0$ term, as we did in the previous section for the strongly disordered case. Here, our justification is that the spectrum has many mini-gaps, so that the contribution of states away from the Fermi energy can be neglected. There are two $E=0$ states, one for each of the two sublattices. Considering the state on the even sites, for example, the same transfer matrix calculation used in the preceding section for the strongly disordered chain holds, so that $\psi(2m)$ has the form shown in \eqref{eq.strongdis}. Specializing to the Fibonacci chain, one can show that the wave function can be expressed in terms of a so-called height function \cite{mace_critical_2017} . The wave function amplitudes for sites on the even sublattice, $i=2m$, can be written as follows
\begin{eqnarray}
\psi(2m) = const (-1)^m \exp(\kappa h(2m)), \label{eq.fiboe0}
\end{eqnarray}
where $\kappa=\log(t_A/t_B)$. The height function $h$, which depends solely on the geometry, can be computed for a given sequence of hopping amplitudes using the following relations for the height changes, which can take three values depending on the value of the hopping amplitudes between the two sites,
\begin{align}
    \delta h(2m) =\begin{cases}
    0 & \text{if } t_{2m-1}=t_{2m}=t_A\\
    -1 & \text{if } t_{2m-1}=t_A, t_{2m}=t_B\\
    1  &\text{if } t_{2m-1}=t_B, t_{2m}=t_A
    \end{cases}
\end{align}
where $\delta h(2m)=h(2m)-h(2m-2)$.
 A similar structure holds for the state on the odd sublattice. 
 
 To proceed, one next uses the renormalization transformation of Fibonacci chains, which relates a given chain to the next generation, to write a recursion relation for the height function. From this relation, one can deduce that for sufficiently long chains the distribution of $h$-values must tend to a Gaussian of width proportional to $\ln L$. The multi-fractal scaling properties of the $E=0$ state can be deduced from the distribution of $h$. In particular, 
for this critical state all the generalized exponents describing its spatial characteristics have been computed exactly. The analysis shows that heights follow a gaussian distribution with the variance given by \cite{mace_critical_2017}
\begin{eqnarray}
\langle h^2\rangle - \langle h\rangle^2 = \frac{1}{\sqrt{5}}\frac{\log(L)}{\log(\tau)}
\label{eq.heightvar}
\end{eqnarray}

Returning to the proximity effect, the OP at the midpoint is determined by the wave functions $u$ and $v$ at the midpoint of the chain which both have the form given by \eqref{eq.fiboe0}, differing only in the values of the prefactor. The changes of values $\Delta_{mid}$ result from the phason flips that occur when the parameter $\phi$ is varied. From \eqref{eq.fiboe0} and \eqref{eq.heightvar} the resulting distribution of $\Delta_{mid}$ must therefore be log-normal. This can be seen in Fig.~\ref{fig:FC-hist}. According to \eqref{eq.fiboe0} the width of this distribution should increase with the strength of quasiperiodic modulation as $\ln(t_B/t_A)$. 

Although both distributions are log-normal, there is a  significant difference between the $L$ dependence of the widths in the Fibonacci chain (FC) as compared to the strongly disordered chain.  For the FC, the width of the distribution grows only logarithmically, much much more slowly than in the random case. This is indeed seen numerically as shown in the inset of Fig. \ref{fig:FC-hist}. In this regard as in many others, the properties of the quasicrystal are intermediate between those of the weakly disordered and strongly disordered chains.

\section{Conclusion}\label{sec:Conclusion}

We have examined the proximity effect in inhomogeneous normal wires coupled to a superconductor at $T=0$, focusing on three important situations : I) when the N component is a weakly disordered crystal, in which wave functions are extended and perturbative calculations can be performed, II) when the disorder is strong enough such that the localization lengths are smaller than the sample size, and III) when states are critical, as in the off-diagonal Fibonacci tight-binding model. Our main results are summarized in table \ref{tab:summary}. We find, firstly, that the \emph{typical} value of the OP has a power law decay as one moves away from the interface when states are extended or critical. On the other hand, for strongly disordered systems, where states are localized, the typical OP decays faster, in the present case as a stretched exponential. For arbitrary positions of the Fermi energy, away from the special point $E=0$ we expect that a regular exponential decay should be observed. 

\begin{table}
\footnotesize\centering
\setlength{\abovecaptionskip}{10pt plus 3pt minus 2pt}
\caption{A summary of the characteristics of the distribution function for  $\Delta_{mid}$ when the electronic states in the normal side are i) (quasi)-extended, as in periodic and weakly disordered systems, ii) critical, as in quasicrystals, iii) localized, as in strongly disordered systems. $\bar\Delta$ and $\Delta_t$ refer to the average (arithmetic mean) and typical (geometric mean) values of the distribution.}\label{tab:summary}
\begin{tabular}{|p{2.25cm}||l|l|l|} \hline
Electronic states                        & $\braket{\Delta_{mid}}$   & $P(\Delta_{mid})$                                             & Variance \\ \hline \hline
Extended/Quasi-extended (Periodic/Weak Anderson model) &  $1/L$                      &  $e^{-\frac{(\Delta-\bar{\Delta})^2}{2\sigma}}$             &  $\sigma^2(\Delta_{mid}) \propto 1/L$             \\                    \hline
Critical (Fibonacci chain)               &  $1/L^\alpha$               &  $\frac{1}{\Delta} e^{-\frac{(\ln\Delta-\ln{\Delta_t})^2}{2\sigma}}$      &   $\sigma^2(\ln\Delta_{mid}) \propto \ln L$              \\             \hline
Localized (Strong Anderson model)              &  $e^{-\sqrt{L/\zeta}}$      &  $\frac{1}{\Delta} e^{-\frac{(\ln\Delta-\ln{\Delta_t})^2}{2\sigma}}$         &    $\sigma^2(\ln\Delta_{mid}) \propto L$              \\     \hline 
\end{tabular}

\end{table}

The OP fluctuations in such systems are large. We have computed the distribution function of OP values and shown that they have gaussian or log-normal shapes. We have presented arguments to explain the forms of the distributions and their scaling as a function of sample size and disorder strength for each of the cases considered.

 In the one-dimensional models we considered there are no mobility edges. However, one can speculate that our results are more generally applicable in other models where there are mobility edges. In that case the typical values and the fluctuations of the OP would depend on the spatial characteristics of the states close to the Fermi level. Lastly, we note that although results reported here were obtained for the 1D case, qualitatively similar results  could be expected for models in higher dimensions. This is left for future studies. 
 
 \section{Acknowledgements}
 The authors acknowledge helpful comments from Nandini Trivedi. G.R. dedicates this work to the memory of Hritik Sampat: your kindness and patience led me and many down the path of physics.

%

\bibliographystyle{apsrev4-1}

\appendix

\end{document}